\documentstyle[12pt,epsfig]{article}
\begin{document}
\title{Survival Analysis, Master Equation, Efficient Simulation of
Path-Related Quantities, and Hidden State Concept of Transitions}
\author{Dirk Helbing}
\maketitle
\section{Introduction}
This paper presents and derives the interrelations between 
survival analysis and master equation. Both have important applications in the
social sciences and other scientific fields treating {\em stochastic systems}.
However, since they focus on different aspects of modeling, it is not yet
generally known that they are closely related to each other.
\par
{\em Survival analysis} deals with modeling the transitions between succeeding
states of a system. Questions related with this are the {\em timing,} 
{\em spacing,} and {\em sequencing} 
of the states of a time series. 
Survival analysis tries to fit and understand
the distribution of these quantities in terms of the functional form of
the {\em hazard rates} which are responsible for the investigated transitions.
The parameters specifying the concrete functional form of the hazard rates
are normally estimated from empirical data
by means of the {\em maximum-} or {\em partial-likelihood method}.
\par
Once the hazard rates are empirically known or 
hypothetically specified, one can carry out microsimulations of 
corresponding time series by means of the {\em Monte-Carlo simulation
method}. This allows the prognosis and investigation of the characteristics
and frequency of time series. 
\par
However, if one is interested in {\em cross-sectional data} connected with
the stochastic process under consideration, one needs to know the temporal 
evolution of the {\em distribution of states}. This can be obtained 
by simulation of the associated 
{\em master equation,} which only presupposes that the initial distribution
and the hazard rates are given. 
In other words: The master equation is 
suitable for the calculation and prognosis
of {\em cross-sectional data} which are related with the longitudinal 
life-table data used for survival analysis. 
\par
In addition, some new formulas are introduced which allow the determination
of path-related (i.e. longitudinal)
quantities like the {\em occurence probability}, the {\em occurence time
distribution}, or the {\em effective cumulative life-time distribution} 
of a certain sequencing of states {\em (path)}. 
These can be efficiently evaluated 
with a recently developed simulation tool (EPIS) which also provides a
new solution method of the master equation (the {\em contracted path integral
solution}). In contrast, a calculation on the basis of time-series data
would require an extreme computational effort.
\par
The effective cumulative life-time distribution
facilitates the formulation of a {\em hidden state concept
of behavioral changes} which allows an interpretation of the respective
time-dependence of hazard rates. Hidden states represent states which are
either not phenomenological distinguishable from other states, not externally
measurable, or simply not detected. They could, for example,
reflect the psychological and mental stages preceding a concrete action,
the individual predisposition, or the internal attitude towards
a certain behavior.

\section{Survival Analysis} \label{SurV}

\subsection{Preliminaries}

Survival analysis deals with the investigation of the {\em timing}, 
{\em spacing,} and {\em sequencing} of {\em longitudinal 
time series (panel studies),} e.g. life-table data
(Blossfeld et al. 1986; Kalbfleisch/Prentice 1980; Elandt-Johnson/Johnson
1980; Diekmann/Mitter 1984; Tuma/Hannan 1984; Cox/Oakes 1984; Lancaster
1990; Courgeau/Leli\`{e}vre 1992).
These time series have the general form
\begin{equation}
 Y(t) = Y_n \qquad \mbox{for} \qquad T_{n} \le t < T_{n+1} \, ,
\end{equation}
where $T_n$ is the variable of the
{\em waiting time} after which the {\em state} $Y_n$ is occupied.
The state $Y_n$ remains occupied for a time period 
$T_{n} \le t < T_{n+1}$ ending directly before the next {\em event}
$(Y_{n+1},T_{n+1})$. Two succeeding events $(Y_{n-1},T_{n-1})$ and
$(Y_n,T_n)$ determine the $n$th {\em episode} 
$((Y_{n-1},T_{n-1}),(Y_{n},T_{n}))$ of a time series.
\par
The state $Y_n$ stems out of a {\em set ${\cal M}_n$ of possible
outcomes} which are often called {\em risks}. Without loss of 
generality\footnote{One can simply define ${\cal M} := 
\bigcup {\cal M}_n$, where ``$\bigcup$'' symbolizes the union of sets.
If, for a given event history $H_{n-1}$, 
no transitions take place to state $j \in {\cal M}$ one must
only set the corresponding transition rate equal to zero, 
e.g. for $j = i_{n-1}$.}
we can assume that this set does not depend on the respective
episode $n$ which implies ${\cal M}_n = {\cal M}$. Furthermore, 
${\cal M}$ is normally a {\em discrete} set so that
\begin{equation}
 Y_n \in {\cal M} = \{ 1, 2, \dots, i, \dots , N \} \, . 
\end{equation}
The elements $i$ of this set represent the possible outcomes. Examples for
${\cal M}$ are:
\begin{equation}
{\cal M} = \{ \mbox{single, married, widowed, divorced} \}
\end{equation} 
or
\begin{equation}
{\cal M} = \{ \mbox{school, army, training, studies, employment, 
 unemployment, retirement} \} \, .
\end{equation} 
\par
Of course, the sequence of waiting times 
$T_n = t_n$ {\em (`timing')} and often also the
sequence of states $Y_n = i_n$ {\em (`sequencing')} 
vary individually so that the actual time series
\begin{equation}
 Y_\alpha(t) = i_n \qquad \mbox{for} \qquad t_{n} \le t < t_{n+1} 
\end{equation}
can only be described by a {\em stochastic process}
\begin{equation}
 (Y,T) := \{ (Y_n,T_n): n = 1, 2, \dots \} \, .
\end{equation}
The subscript $\alpha \in \{1,2, \dots, A\}$ 
distinguishes the different individuals, systems, or realizations
to which the respective time series belong.
\par
Now, the {\em spacing} of a time series can be defined as the sequence
of {\em life times (survival times, failure times)}
\begin{equation}
 V_n := T_n - T_{n-1} \, , \qquad k = 1, 2, \dots \, .
\end{equation}
Additionally, we introduce the so-called {\em event history} of a
state $Y_n = i_n$ which has been occupied at time $T_n = t_n$ by
\begin{equation}
 H_{n-1} := \{ t_0, i_0;t_1, i_1, {\bf x}_1; \dots ;
 t_{n-1}, i_{n-1}, {\bf x}_{n-1} \} \, .
\end{equation}
Here, the {\em vector of covariates} ${\bf x}_k = (x_{k,1}, \dots, x_{k,M})$ 
comprises the different factors $x_{k,m}$ like education or sex which may influence
the {\em transition} (i.e. change) from the $(k-1)$st 
state $Y_{k-1} = i_{k-1}$ to the 
$k$th state $Y_k = i_k$. 
\par
The following figure illustrates the terms which were introduced above:
\par\begin{figure}[htbp]
\unitlength0.75cm
\begin{center}
\begin{picture}(20,8.3)(1,0)
\put(1.8,3){\vector(1,0){13}}
\put(2,2.8){\vector(0,1){4.5}}
\put(2,3.5){\dashbox{0.2}(12,0)}
\put(2,4.5){\dashbox{0.2}(12,0)}
\put(2,5.5){\dashbox{0.2}(12,0)}
\put(2,6.5){\dashbox{0.2}(12,0)}
\put(14,2.8){\vector(0,1){4.5}}
\thicklines
\put(2,3.5){\line(1,0){4}}
\put(6,3.5){\vector(0,1){2}}
\put(6,5.5){\line(1,0){3}}
\put(9,5.5){\vector(0,-1){1}}
\put(9,4.5){\line(1,0){5}}
\put(14,4.5){\vector(0,1){2}}
\thinlines
\put(2,0.8){\line(0,1){0.4}}
\put(2,1){\vector(1,0){4}}
\put(6,1){\vector(-1,0){4}}
\put(6,0.8){\line(0,1){0.4}}
\put(6,1){\vector(1,0){3}}
\put(9,1){\vector(-1,0){3}}
\put(9,0.8){\line(0,1){0.4}}
\put(9,1){\vector(1,0){5}}
\put(14,1){\vector(-1,0){5}}
\put(14,0.8){\line(0,1){0.4}}
\put(2,8){\makebox(0,1){\small (Possible)}}
\put(2,7.2){\makebox(0,1){\small States $i$}} 
\put(1.2,3.5){\makebox(0,0){\small 1}}
\put(1.8,3.5){\line(1,0){0.2}}
\put(1.2,4.5){\makebox(0,0){\small 2}}
\put(1.8,4.5){\line(1,0){0.2}}
\put(1.2,5.5){\makebox(0,0){\small 3}}
\put(1.8,5.5){\line(1,0){0.2}}
\put(1.2,6.5){\makebox(0,0){\small 4}}
\put(1.8,6.5){\line(1,0){0.2}}
\put(2,2.2){\makebox(0,0){\small $t_0$}}
\put(2,2.8){\line(0,1){0.2}}
\put(6,2.2){\makebox(0,0){\small $t_1$}}
\put(6,2.8){\line(0,1){0.2}}
\put(9,2.2){\makebox(0,0){\small $t_2$}}
\put(9,2.8){\line(0,1){0.2}}
\put(14,2.2){\makebox(0,0){\small $t_3$}}
\put(14,2.8){\line(0,1){0.2}}
\put(4,1.5){\makebox(0,0){\small $v_1 = t_1 - t_0$}}
\put(7.5,1.5){\makebox(0,0){\small $v_2 = t_2 - t_1$}}
\put(11.5,1.5){\makebox(0,0){\small $v_3 = t_3 - t_2$}}
\put(4,0.4){\makebox(0,0){\small 1}}
\put(7.5,0.4){\makebox(0,0){\small 2}}
\put(11.5,0.4){\makebox(0,0){\small 3}}
\put(15,3){\makebox(0,0)[l]{\small Time $t$}}
\put(15,2.2){\makebox(0,0)[l]{\small Waiting Times $T_n$}}
\put(15,1.4){\makebox(0,0)[l]{\small Life Times $V_n = T_n - T_{n-1}$}}
\put(15,0.4){\makebox(0,0)[l]{\small Episodes $n$}}
\put(15,8){\makebox(0,1){\small Occupied}}
\put(15,7.2){\makebox(0,1){\small States $Y_n = i_n$}}
\put(14.8,3.5){\makebox(0,0){\small $i_0$}}
\put(14,3.5){\line(1,0){0.2}}
\put(14.8,4.5){\makebox(0,0){\small $i_2$}}
\put(14,4.5){\line(1,0){0.2}}
\put(14.8,5.5){\makebox(0,0){\small $i_1$}}
\put(14,5.5){\line(1,0){0.2}}
\put(14.8,6.5){\makebox(0,0){\small $i_4$}}
\put(14,6.5){\line(1,0){0.2}}
\end{picture}
\end{center}
{\small Figure 1: Example illustrating a time series with $N=4$ competing risks
and 3 episodes which
are given by the sequence of events $(1,t_0) \rightarrow (3,t_1) 
\rightarrow (2,t_2) \rightarrow (4,t_3)$.}
\end{figure}

\subsection{Central Concepts}

We now come to the definition of some central concepts of survival
analysis. In view of the following discussion, 
we restrict ourselves to the special case that
\begin{itemize}
\item[1.] only the last state $Y_{n-1} = i_{n-1}$ 
of the event history $H_{n-1}$
has an influence on the $n$th transition {\em (Markov case),}
\item[2.] the transitions and the vector of covariates
${\bf x}_n$ are independent of the respective episode $n$.
\end{itemize}
The second assumption implies that the vector of covariates ${\bf x}$ is
time-independent during the survey. This means that each individual or system
$\alpha$ is characterized by a fixed value of ${\bf x}$ and that different
vectors of covariates distinguish different {\em cohorts (subpopulations)}.
\par
Then, the so-called {\em hazard rates (transition rates)} 
$\lambda_{\bf x}(t;j|i)$ of subpopulation ${\bf x}$ are defined
as the probability $P_{\bf x}(T_n < t + \Delta t , Y_n = j | T_n \ge t,
Y_{n-1}=i)$\footnote{Instead of $P_{\bf x}(T_n < t + \Delta t , 
 \mbox{$Y_n = j |$} T_n \ge t, Y_{n-1}=i )$ one sometimes writes 
$P_{\bf x}(t \le T_n < t + \Delta t, \mbox{$Y_n = j |$} T_n \ge t,
 Y_{n-1}=i )$. However, the expression $t \le T_n < t + \Delta t$
is unnecessary complicated since $t \le T_n$ is already presupposed by the
condition $T_n \ge t$ in the second part of the argument of $P_{\bf x}$.} 
per unit time $\Delta t > 0$ to change
into state $Y_n = j$ up to time $t + \Delta t$ on the conditions
that the $n$th transition did not happen before time $t$ and the preceding
state was $Y_{n-1} = i$:
\begin{equation}
 \lambda_{\bf x}(t;j|i) := \lim_{\Delta t \rightarrow 0}
 \frac{1}{\Delta t} P_{\bf x}(T_n < t + \Delta t , Y_n = j | T_n \ge t,
 Y_{n-1}=i )  \, .
\label{rates}
\end{equation}
\par
Moreover, the {\em survivor function (survival function)} 
$S_{\bf x}^n(t|i)$ of cohort ${\bf x}$ is defined by the 
probability $P_{\bf x}(T_n \ge t|Y_{n-1}=i )$ that the
$n$th transition does not take place before time $T_n$, given that the 
preceding state was $Y_{n-1} = i$:
\begin{equation}
 S_{\bf x}^n(t|i) :=  P_{\bf x}(T_n \ge t|Y_{n-1}=i ) \, .
\end{equation}
One can find the relation
\begin{equation}
 S_{\bf x}^n(t|i) = 1 - F_{\bf x}^n(t|i ) 
\label{SF}
\end{equation}
with regard to the {\em cumulative distribution function 
(life-time distribution, failure distribution, duration distribution
function)}
\begin{equation}
 F_{\bf x}^n(t|i) := P_{\bf x}(T_n < t|Y_{n-1}=i ) 
\label{lifetime}
\end{equation}
of subpopulation ${\bf x}$. The latter describes the probability that
the $n$th transition happens before time $t$, given that the preceding
state was $Y_{n-1} = i$. 
If we are interested in the probability that the $n$th transition not only
happens before time $t$ but additionally leads to state $Y_n = j \ne i$, 
we need the quantity
\begin{equation}
 F_{\bf x}^n(t;j|i) := P_{\bf x}(T_n < t,Y_n = j|Y_{n-1}=i ) 
\label{need}
\end{equation}
which fulfils
\begin{equation}
 \sum_{j=1 \atop (j\ne i)}^N \!\! F_{\bf x}^n(t;j|i) = 
 P_{\bf x}(T_n < t|Y_{n-1}=i ) =  F_{\bf x}^n(t|i) \, .
\label{need2}
\end{equation}
\par
Finally we define the {\em probability density function
(failure time (sub)density function for failure type $j$)} by
\begin{eqnarray}
 f_{\bf x}^n(t;j|i) &:=& \lim_{\Delta t \rightarrow 0}
 \frac{1}{\Delta t} P_{\bf x}(t \le T_n < t + \Delta t , Y_n = j |
 Y_{n-1}=i ) \nonumber \\
 &=& \lim_{\Delta t \rightarrow 0} 
 \frac{P_{\bf x}(T_n < t + \Delta t, Y_n = j |Y_{n-1}=i ) 
 - P_{\bf x}(T_n < t, Y_n = j |Y_{n-1}=i ) }{\Delta t} \nonumber \\
 &=& \frac{d}{dt} F_{\bf x}^n(t;j|i) \, .
\label{one}
\end{eqnarray}
(\ref{rates}) to (\ref{SF}), and (\ref{one}) imply the relations
\begin{equation}
 \sum_{j=1 \atop (j \ne i)}^N \!\! f_{\bf x}^n(t;j|i)
 = \frac{d}{dt} F_{\bf x}^n(t|i) 
 = - \frac{d}{dt} S_{\bf x}^n(t|i) 
\label{three}
\end{equation}
and
\begin{eqnarray}
 & & \hspace*{-2cm} \lambda_{\bf x}(t;j|i) S_{\bf x}^n(t|i) 
 = \lambda_{\bf x}(t;j|i) P_{\bf x}(T_n \ge t|Y_{n-1}=i )
 \nonumber \\
 &=& \lim_{\Delta t \rightarrow 0}
 \frac{1}{\Delta t} P_{\bf x}(T_n < t + \Delta t , Y_n = j | T_n \ge t,
 Y_{n-1}=i ) P_{\bf x}(T_n \ge t|Y_{n-1}=i ) \nonumber \\
 &=& \lim_{\Delta t \rightarrow 0}
 \frac{1}{\Delta t} P_{\bf x}(T_n < t + \Delta t , Y_n = j ,  T_n \ge t |
 Y_{n-1}=i ) \nonumber \\
 &=& f_{\bf x}^n(t;j|i) \, .
\label{two}
\end{eqnarray}
Equations (\ref{three}) and (\ref{two}) yield the 
differential equation
\begin{equation}
 \sum_{j=1 \atop (j \ne i)}^N \!\!
 \lambda_{\bf x}(t;j|i) S_{\bf x}^n(t|i) 
 = - \frac{d}{dt} S_{\bf x}^n(t|i) 
\end{equation}
which has the important solution
\begin{equation}
 S_{\bf x}^n(t|i) = \exp \left[ - \!\int\limits_{t_{n-1}}^t \!
 dt' \, \lambda_{\bf x}(t'|i) \right] 
 = \prod_{j=1 \atop (j \ne i)}^N \exp \left[ - \!\int\limits_{t_{n-1}}^t \!
 dt' \, \lambda_{\bf x}(t';j|i) \right]
\label{eq}
\end{equation}
with the {\em overall departure rate (overall hazard rate, overall failure rate)}
\begin{equation}
 \lambda_{\bf x}(t|i) := \! \sum_{j=1 \atop (j \ne i)}^N
 \lambda_{\bf x}(t;j|i) \, .
\label{whynot1}
\end{equation}
That is, survivor functions are determined by exponential 
relations which only depend on the hazard rates.

\section{The Master Equation Technique} \label{MasteR}
 
\subsection{Derivation of the Master Equation}

We will now derive a system of differential equations 
for the evolution of
the {\em probability distribution} $P_{\bf x}(j,t)$ 
of states $j$ with time $t$ which is
associated with the above described stochastic process. 
(Each subpopulation {\bf x} obeys its own system of equations.)
For this purpose
we apply two relations from probability theory:
\begin{equation}
 \sum_{i=1}^N P_{\bf x}(i , t'|j, t) = 1
\label{eins}
\end{equation}
and
\begin{equation}
 P_{\bf x}(j,t') = \sum_{i=1}^N 
 P_{\bf x}(j , t'|i, t) P_{\bf x}(i,t) \, .
\label{zwei}
\end{equation}
Here, $P_{\bf x}(j, t'|i,t)$ denotes the probability that
we have state $j$ at time $t'$ given that we had state $i$ at time $t$. 
With (\ref{eins}) and (\ref{zwei}) we get
\begin{eqnarray}
 & & \hspace*{-8mm} \lim_{\Delta t \rightarrow 0} 
 \frac{P_{\bf x}(j,t+\Delta t) - P_{\bf x}(j,t)}{\Delta t} \nonumber \\
 &=& \lim_{\Delta t \rightarrow 0} \frac{1}{\Delta t} \left[ \left(
 \sum_{i=1}^N P_{\bf x}(j , t + \Delta t|i, t)  
 P_{\bf x}(i,t) \right) - 
 \left( \sum_{i=1}^N P_{\bf x}(i , t + \Delta t|j, t) \right)
 P_{\bf x}(j,t) \right] \\
 &=& \sum_{i=1 \atop (i \ne j)}^N \left[ \left( \lim_{\Delta t \rightarrow 0}  
 \frac{1}{\Delta t}
 P_{\bf x}(j , t + \Delta t|i, t) \right) P_{\bf x}(i,t) 
 -  \left( \lim_{\Delta t \rightarrow 0}
 \frac{1}{\Delta t} P_{\bf x}(i , t + \Delta t|j, t) 
 \right)  P_{\bf x}(j,t) \right] \, . \nonumber
\end{eqnarray}
Due to 
\begin{equation}
 \lim_{\Delta t \rightarrow 0} \frac{1}{\Delta t} 
 P_{\bf x} (j, t + \Delta t | i, t) 
 = \lim_{\Delta t \rightarrow 0} \frac{1}{\Delta t} 
 P_{\bf x}(T_n < t + \Delta t , Y_n = j | T_n \ge t, Y_{n-1}=i )
\end{equation}
and (\ref{rates}) we finally obtain the differential equation
\begin{equation}
  \frac{d}{dt} P_{\bf x}(j,t) =
 \sum_{i=1 \atop (i \ne j)}^N \Big[ \lambda_{\bf x}(t;j|i) P_{\bf x}(i,t)
 - \lambda_{\bf x}(t;i|j) P_{\bf x}(j,t) \Big]
\label{masteR}
\end{equation}
which is called the {\em master equation}
(Haken 1983; Weidlich/Haag 1983; Helbing 1995). Chiang (1968: 116ff),
Lancaster (1990: pp. 109ff), and Courgeau/Leli\`{e}vre (1992: 40) 
presented related considerations for {\em continuous-time Markov processes}
leading to the {\em Kolmogorov differential equation} 
\begin{equation}
 \frac{d}{dt} P_{\bf x}(j,t|i_0,t_0) = 
 \sum_{i=1}^N \lambda_{\bf x}(t;j|i) P_{\bf x}(i,t|i_0,t_0) 
 \quad \mbox{with} \quad \lambda_{\bf x}(t;j|j) := - 
 \sum_{i=1\atop (i\ne j)}^N \lambda_{\bf x}(t;i|j) \, .
\label{Kolm}
\end{equation}
As known from the theory of stochastic processes, the master equation 
can be obtained from (\ref{Kolm}) by multiplication with the
initial distribution of states $P_{\bf x}(i_0,t_0)$ and subsequent summation 
over the initial states $i_0$ because of 
$P_{\bf x}(j,t|i_0,t_0)P_{\bf x}(i_0,t_0) = P_{\bf x}(j,t;i_0,t_0)
= P_{\bf x}(i_0,t_0|j,t) P_{\bf x}(j,t)$ and (\ref{eins}). 
\par
According to the master equation,
the temporal change of the probability $P_{\bf x}(j,t)$ to have
state $j$ at time $t$ is given by the sum of the {\em effective}
transition rates $\lambda_{\bf x}(t;j\leftarrow i)$ 
from other states $i$ to state $j$ minus
the sum of the effective transition rates 
$\lambda_{\bf x}(t;i\leftarrow j)$ from state $j$ to other states $i$.
The {\em effective transition rate} 
$\lambda_{\bf x}(t;j\leftarrow i) := \lambda_{\bf x}(t;j|i) P_{\bf x}(i,t)$
from state $i$ to state $j$ is obviously the product of the transition rate
$\lambda_{\bf x}(t;j|i)$ {\em given} that the indiviual or system 
under consideration is in state $i$ times the 
probability $P_{\bf x}(i,t)$ that he/she/it
is {\em actually} in state $i$.
\par
In the following we will assume the special case that we have the dependence
\begin{equation}
 \lambda_{\bf x}(t;j|i) := \lambda_{\bf x}^0(t) w_{\bf x}(j|i) 
\label{prod}
\end{equation}
of the transition rates. Often one even restricts oneself to the
{\em proportional hazard rate model (proportional hazards model)}
\begin{equation}
 \lambda_{\bf x}(t;j|i) := \lambda_0(t) 
 \exp (\mbox{\boldmath $\beta$}_{\!ji} {\bf x})
\end{equation}
so that $\lambda_{\bf x}^0(t) = \lambda_0(t)$
with the {\em baseline hazard function} $\lambda_0(t)$ and
$w_{\bf x}(j|i) = \exp (\mbox{\boldmath $\beta$}_{\!ji} {\bf x})$.
The optimal {\em parameter vectors} $\mbox{\boldmath $\beta$}_{\!ji}$ 
can be estimated from life-table data via the {\em partial likelihood method}
proposed by Cox (1975) (see also Blossfeld et al. 1986; Kalbfleisch/Prentice 
1980; Elandt-Johnson/Johnson 1980; Diekmann/Mitter 1984; Tuma/Hannan 1984; 
Cox/Oakes 1984; Lancaster 1990; Courgeau/Leli\`{e}vre 1992). 
\par
For reasons of simplicity we utilize relation (\ref{prod}) to
introduce the subpopulation-specific times
\begin{equation}
 \tau_{\bf x}(t) := \int\limits_{t_0}^t dt'\, \lambda_{\bf x}^0(t')
 \qquad \mbox{corresponding to} \qquad
 \frac{d\tau_{\bf x}}{dt} = \lambda_{\bf x}^0(t) \, .
\end{equation}
This implies, for example,
\begin{equation}
 \frac{d}{dt} P_{\bf x}(j,t) = 
 \frac{d}{dt} P_{\bf x}[j,\tau_{\bf x}(t)] = \left( \frac{d}{d\tau_{\bf x}}
 P_{\bf x}(j,\tau_{\bf x}) \right)
 \frac{d\tau_{\bf x}}{dt} = \left( \frac{d}{d\tau_{\bf x}}
 P_{\bf x}(j,\tau_{\bf x}) \right) \lambda_{\bf x}^0(t) \, .
\end{equation}
As a consequence, we can bring the master equation into the form
\begin{equation}
 \frac{d}{d\tau_{\bf x}}P_{\bf x}(j,\tau_{\bf x}) = 
 \sum_{i=1 \atop ( i \ne j)}^N \Big[ w_{\bf x}(j|i) P_{\bf x}(i,\tau_{\bf x})
 - w_{\bf x}(i|j) P_{\bf x}(j,\tau_{\bf x}) \Big] 
\end{equation}
with {\em time-independent} transition rates
\begin{equation}
 \lambda_{\bf x}(\tau_{\bf x};j|i) = w_{\bf x}(j|i) \, . 
\end{equation}
Analogous simplifications are found for the other quantities:
\begin{equation}
  \lambda_{\bf x}(\tau_{\bf x}|i) = w_{\bf x}(i)
 := \sum_{j=1 \atop (j \ne i)}^N w_{\bf x}(j|i) \, ,
\end{equation}
\begin{equation}
 S_{\bf x}^n(\tau_{\bf x}|i) = \exp \left[ - w_{\bf x}(i) 
 (\tau - \tau_{{\bf x}, n-1}) \right] = 
 1 - F_{\bf x}^n(\tau_{\bf x}|i) = 1 - \!
 \sum_{j=1 \atop (j \ne i)}^N F_{\bf x}^n(\tau_{\bf x};j|i) \, ,
\label{accord}
\end{equation}
and
\begin{equation}
 f_{\bf x}^n(\tau_{\bf x};j|i) = w_{\bf x}(j|i) S_{\bf x}^n(\tau_{\bf x}|i) 
 = \frac{d}{d\tau_{\bf x}} F_{\bf x}^n(\tau_{\bf x};j|i) \, , 
\label{exactly}
\end{equation}
where we have used the convention $\tau_{{\bf x},n} := \tau_{\bf x}(t_n)$.
Since the subscript ${\bf x}$ for the subpopulation (cohort) 
is arbitrary but fixed (time-independent) we will omit it in the following
which makes the mathematical formulas easier to read.

\subsection{Simulations with the Master Equation} \label{MasterSim}

The master equation has proved to be a very powerful and quite flexible
tool for the description and simulation of stochastically behaving systems
which are subject to inherent or external random influences ('fluctuations').
It has got numerous applications in physics, chemistry, biology, economics,
and the social sciences (Weidlich/Haag 1983; Troitzsch 1990;
Weidlich 1991; Helbing 1995). The master equation can normally not be 
analytically solved. However, since the master equation has the form of 
a linear system of ordinary differential equations,
it can be easily simulated by means of the usual 
numerical integration algorithms (Press et al. 1992: Chap. 16).
\par
With today's computers these simulations are very fast, so that even systems
with some dozens of variables can be solved within a few minutes.
Therefore, the master equation technique facilitates an efficient evaluation
of all {\em cross-sectional quantities} related with stochastic processes.
This includes the temporal course of the distribution of states $P(j,\tau)$,
its maxima, means value, variance, etc. Whereas the mean value represents
the average behavior of a huge number of time series, the course of the
most probable time series will often (but not always) 
be close to the maxima of the distribution of states, and the variance is a
measure for the grade of variation between different time series. In the
case of a multimodal distribution of states, the mean value can considerably
differ from the most probable time series.
\par
In summary, simulations of the master equation are suitable for 
{\em scenario techniques}, since they allow
\begin{itemize}
\item[1.] to investigate and compare the properties of different
  conceivable stochastic models (which are normally related to different
  functional forms of the hazard rates),
\item[2.] to evaluate the implications of parameter changes
(which may correspond to considered or planned modifications of legal
regulations or social conditions),
\item[3.] to make prognoses of the range of probable future behaviors
of a stochastic system.
\end{itemize}

\section{Simulation of Path-Related Quantities}

\subsection{Microsimulations with the Monte-Carlo Technique} \label{MC}

Although the master equation technique is very versatile, it does not allow
investigations which are directly related with the {\em longitudinal
time series.} However, the simulation of these is desirable for a number
of reasons:
\begin{itemize}
\item[1.] The single time series give an impression of the possible
  time-dependent behavior of individual systems. If the variation among
  different time series is small, their predictive value is large. However, if
  the state changes of the individual time series are large and frequent, they
  are normally not representative of a system's behavior. Note that, by
  means of the time-dependent distribution of states (i.e. by simulation of
  the master equation alone), we cannot always distinguish between a strongly
  varying (e.g. oscillatory) system behavior and a ``smooth'' one.
\item[2.] One can often distinguish {\em desirable} and {\em undesired} 
(maybe catastrophic) system states. In such cases we may be interested in
the {\em occurence probability} with which the system under consideration 
up to time $t$
takes a {\em desired time series} (containing desirable states only).
This is relevant for {\em prognoses} 
as well as for the {\em controllability} of technical systems. 
\item[3.] Sometimes one also likes to know the 
{\em occurence time distribution} of a certain sequencing of states.
\item[4.] When occupying an undesired state, it may be interesting to
evaluate the expected {\em escape time} until the system reaches, for the
first time, one of the desired states.
\end{itemize}
All these quantities can be obtained by evaluating a large number of
stochastic time series. These can
be generated and investigated in {\em microsimulations}
applying the {\em Monte-Carlo simulation method} 
(cf. Binder 1979: Chap. 1.3.1). 
\par
The Monte-Carlo technique 
utilizes the fact that a system under consideration stays
in an occupied state $i_{n-1}$ for a time period $\tau_{n} - \tau_{n-1}$ 
which is exponentially
distributed according to (\ref{accord}). Therefore, in Monte-Carlo
simulations the respective life time $\tau_{n} - \tau_{n-1}$ 
is determined via the formula
\begin{equation}
 \tau_{n} - \tau_{n-1} := - \frac{1}{w(i_{n-1})} \ln y_{n} \, ,
\end{equation}
where $y_{n} \in [0,1]$ are uniformly distributed random numbers
(cf. Press et al. 1992: Chap. 7).
After this time period (i.e. at time $\tau_n$) the system 
goes over into one of the other
states $i_{n} \ne i_{n-1}$ with probability
\begin{equation}
 p(i_{n}|i_{n-1}) := \frac{w(i_{n}|i_{n-1})}{\displaystyle
 \sum_{j=1\atop (j \ne i_{n-1})}^N w(j|i_{n-1})}
 = \lim_{\Delta \tau \rightarrow 0}
 \frac{P(i_{n},\tau_{n}+\Delta \tau|i_{n-1},\tau_{n})}{\displaystyle
 \sum_{j=1\atop (j \ne i_{n-1})}^N P(j,\tau_{n}+\Delta \tau|i_{n-1},
 \tau_{n})} \, .
\end{equation}
The realized state $i_n$ is again selected by means of a uniformly distributed
random variable $z_{n} \in [0,1]$: The randomly resulting value $z_{n}$ 
corresponds to an occupation of the state $i_{n}$ for which
\begin{equation}
 Z(i_{n-1}) < z_{n} \le Z(i_{n}) \qquad \mbox{with} \qquad 
 Z(i_n) = P(j \le i_n|i_{n-1}) := \sum_{j=1}^{i_{n}} p(j|i_{n-1}) 
\end{equation}
is fulfilled. The initial state $i_0$ is usually given, but 
alternatively it can also be randomly chosen
in a similar way.
\par
Time series which are generated by means of the above outlined method have the
meaning of possible realizations of the stochastic process under 
consideration. They can be investigated in different ways. For
illustrative reasons,
let us assume to have generated $A$ different time series $Y_\alpha(\tau)$.
If $n_j$ is the number of systems found in state $j$ at time $\tau$, the
distribution of states corresponds to
\begin{equation}
 P(j,\tau) = \lim_{A \rightarrow \infty} \frac{n_j(\tau)}{A} \qquad
 \mbox{since} \qquad \sum_{j=1}^N n_j(\tau) = A \, .
\end{equation}
For finite $A$, the probability $P(n_1,\dots,n_N;\tau)$ to find the
{\em occupation numbers} $n_j$ at time $\tau$ is given by the
{\em multinomial distribution}
\begin{equation}
 P(n_1,\dots,n_N;\tau) = \frac{A!}{n_1! \dots n_N!}
 \prod_{j=1}^N P(j,\tau)^{n_j} 
\label{oder}
\end{equation}
(cf. Helbing 1995: 70), so that the distribution of states $n_j(\tau)/A$ 
obtained by microsimulations will normally differ from $P(j,\tau)$.
This shows that we would need a tremendous number $A$ of simulation runs
to determine $P(j,\tau)$ from microsimulations. For other quantities
the situation is similar. As a consequence,
the Monte-Carlo technique is rather a brute force than an 
efficient method. It is, however, very suitable for generating some sample
time series for illustrative purposes (which is sufficient for Item 1). The
big advantage of the microsimulation technique is that it usually allows a
relatively simple treatment of systems with a huge number of possible system 
states and/or very complex relations for the transition rates.
Therefore, it is mainly used in situations where the derivation or simulation
of a master equation is too difficult.
\par
Since the microsimulation technique normally requires an extreme 
computational effort, it should only be applied
if there are no suitable alternatives.
As discussed in Section \ref{MasterSim}, the distribution of states 
$P(j,\tau)$ is much easier obtained by simulating the master equation.
In the next section it will be shown that more efficient methods than
the microsimulation technique can also be developed for 
the problems raised in Items 2, 3, and 4. The reason is, that they only
regard the {\em sequencing} of time series up to a certain time $\tau$
but not the {\em waiting times} after which the single state of the
time series are occupied. 

\subsection{Occurence Probabilities and Occurence Times of Paths}

With the results of sections \ref{SurV} and \ref{MasteR}
we can now calculate the
probability $P(H_n,t)$ that we have the event history
\begin{equation}
 H_n := \{ t_0,i_0;t_1,i_1;\dots;t_n,i_n\}
 = \{ \tau_0,i_0;\tau_1,i_1;\dots;\tau_n,i_n\} 
\end{equation}
at time $\tau(t)$. Remembering that $S^k(\tau_k|i_{k-1}) 
= P({\cal T}_k \ge \tau_k|Y_{k-1}=i_{k-1})$ 
with
\begin{equation}
 {\cal T}_k := \tau(T_k) 
\end{equation}
is the probability to stay in state $i_{k-1}$ up to time $\tau_k$ and 
that $w(i_k|i_{k-1})d\tau_k = P(i_k, \tau_k+ d\tau_k | i_{k-1}, \tau_k)$ 
is the probability to change from state $i_{k-1}$ to state $i_k$ between 
times $\tau_k$ and $\tau_k+d \tau_k$, we find
\begin{eqnarray}
 P(H_n,\tau) &=& S^{n+1}(\tau|i_n) w(i_n|i_{n-1})d\tau_n
 \, S^n(\tau_n|i_{n-1}) w(i_{n-1}|i_{n-2})d\tau_{n-1} \nonumber \\
& & \quad \dots w(i_2|i_1)d\tau_2 \, S^2(\tau_2|i_1)
 w(i_1|i_0)d\tau_1 \, S^1(\tau_1|i_0) P(i_0,\tau_0) 
\label{dies}
\end{eqnarray}
(Empacher 1992).
Often, however, one is not interested in the times $\tau_k= \tau(t_k)$ 
at which
the single transitions occur, but only in the sequencing,
i.e. in the {\em path}
\begin{equation}
 {\cal C}_n := i_n \longleftarrow i_{n-1} \longleftarrow \dots \longleftarrow
 i_1 \longleftarrow i_0 
\end{equation}
which the system has taken up to time $\tau(t)$. For example, one could
ask which is the probability that somebody is married the second time
or unemployed the third time at time $\tau$. One could also compare
the probabilities of being married the second time after divorce
or after death of the partner. 
The necessary quantities for answering questions like these can be derived
from formula (\ref{dies}) by integration with respect to $\tau_1, \dots, \tau_n$.
For the probability of having path ${\cal C}_n$ at time $\tau$ we obtain
\begin{equation}
 P({\cal C}_n,\tau) = \sum_{k=0}^n 
 \frac{\mbox{e}^{-w_k \tau}}
 {\displaystyle \prod_{l=0 \atop (w_l\ne w_k)}^n (w_l - w_k)} \, 
 p_{m_k}(w_k,\tau) w({\cal C}_n) P(i_0,t_0) \, .
\label{form}
\end{equation}
Here, $m_k$ is the {\em multiplicity} of the overall departure rate
$w_k := w(i_k)$ in path ${\cal C}_n$,
\begin{equation}
 w({\cal C}_n) = w(i_n \leftarrow i_{n-1} \leftarrow \dots \leftarrow i_0)
 := \left\{
\begin{array}{ll}
1 & \mbox{if $n=0$} \\
{\displaystyle \prod_{l=1}^n w(i_l|i_{l-1})} & \mbox{if $n\ge 1$,}
\end{array}\right.
\end{equation}
\begin{eqnarray}
 p_{m} &=& \frac{(-1)^{m+1}}{m (m - 1)}
 \Bigg( g^{(m-1)} + \sum_{n_1 = 1}^{m - 2}
 \frac{g^{(m-1-n_1)}}{n_1} \Bigg( g^{(n_1)} 
 + \sum_{n_2 = 1}^{n_1 - 1} \frac{g^{(n_1 - n_2)}}{n_2} \nonumber \\
&\times & \Bigg(g^{(n_2)} + \dots + \sum_{n_{n_0} -1 = 1}^{n_{n_0 - 2} - 1}
 \Bigg( \frac{g^{(n_{n_0-2} - n_{n_0 - 1})}}{n_{n_0-1}}
 g^{(n_{n_0 - 1})} \Bigg) \dots \Bigg) \Bigg) \Bigg) 
\label{endlich}
\end{eqnarray}
for $m \ge 2$, $p_m = 1$ for $m=1$, and
\begin{equation}
 g^{(l+1)}(w_k,\tau) :=  \sum_{l=0\atop(w_l \ne w_k)}^n 
 \frac{1}{(w_l - w_k)^{l+1}}  - \tau \delta_{l0} 
\end{equation}
with $\delta_{l0}=1$ if $l=0$ and $\delta_{l0} = 0$ otherwise.
The detailed steps on the way of deriving $P({\cal C}_n,\tau)$
are presented in a paper by Helbing and Molini (1995).
If one restricts to the case that
all overall departure rates $w_k$ are different from each other 
(i.e. to pure {\em birth processes}) the above
relation simplifies to
\begin{equation}
 P({\cal C}_n,\tau) =
 \sum_{k=0}^n \frac{S^1(\tau|i_k)}{\displaystyle
 \prod_{l=0 \atop (l\ne k)}^n [w(i_l) - w(i_k)] } w({\cal C}_n)P(i_0,\tau_0)
 \, , \quad \mbox{where} \quad 
 S^1(\tau|i_k) = \mbox{e}^{- w(i_k) (\tau - \tau_0)} \, .
\label{erg1}
\end{equation} 
The special formula (\ref{erg1}) was already presented by Chiang (1968: 50ff) 
who, however, did not introduce the more general path concept developed above.
\par
The {\em probability density $P(\tau|{\cal C}_n)$ of the occurence times}
$\tau$ (the {\em occurence time distribution})
of path ${\cal C}_n$ can now easily be obtained from (\ref{form}).
It is given by the formula
\begin{equation}
 P(\tau|{\cal C}_n) = \frac{P({\cal C}_n,\tau)}{P({\cal C}_n)} \quad
 \mbox{with} \quad P({\cal C}_n) := \int\limits_{\tau_0}^\infty d\tau \, 
 P({\cal C}_n, \tau) = w({\cal C}_n) P(i_0,\tau_0) 
 \prod_{k=0}^n \frac{1}{w(i_k)}  
\label{occtime}
\end{equation}
(Helbing, 1994, 1995).
From this we can derive the {\em average}
\begin{equation}
 \langle \tau \rangle_{{\cal C}_n} = \int\limits_{\tau_0}^\infty
 d\tau \, \tau P(\tau|{\cal C}_n) = \sum_{k=0}^n \frac{1}{w(i_k)}
\end{equation}
{\em of the occurence times} $\tau$ and their {\em variance}
\begin{equation}
 \Theta_{{\cal C}_n} = \Big\langle ( \tau - \langle \tau \rangle_{{\cal C}_n}
 )^2 \Big\rangle_{{\cal C}_n} = \sum_{k=0}^n \frac{1}{[w(i_k)]^2} 
\end{equation}
(cf. Helbing 1994, 1995). The average occurence times are the basis for the
calculation of the expected {\em escape times}.

\subsection{The Simulation Tool EPIS}

For a determination of the quantities called for by Items 2 and 4 of
Section \ref{MC} we have to sum up over the occurence probabilities or
average occurence times of many paths. In the following, we will discuss
how this can be done in an efficient way with respect to computer memory
and simulation time. For illustrative reasons 
we begin with the socalled {\em contracted 
path integral solution} of the master equation. This reads
\begin{equation}
 P(j,\tau) = \sum_{n=0}^\infty \sum_{{\cal C}_n} P({\cal C}_n,\tau)
\label{erg2}
\end{equation}
with
\begin{equation}
 \sum_{{\cal C}_n} := \!\! \sum_{i_{n-1}=1 \atop (i_{n-1} \ne j)}^N
 \sum_{i_{n-2}=1 \atop (i_{n-2} \ne i_{n-1})}^N \dots
 \sum_{i_0=1 \atop (i_0 \ne i_1)}^N \, ,
\label{sechs}
\end{equation}
since, using $\tau_0 = 0$, the probability $P(j,\tau)$ to find state 
$j$ at time $\tau$ is given as the sum over the occurence
probabilities $P({\cal C}_n,\tau)$ of all paths 
${\cal C}_n$ with an arbitrary length $n$
which lead to state $i_n := j$ within the time interval $\tau$
(Helbing 1994, 1995). 
\par
When (\ref{erg2}) is evaluated numerically, one must 
restrict the summation to a finite number of {\em relevant} paths.
Here, we can utilize the fact that the occurence probability 
$P({\cal C}_n,\tau)$ of a path ${\cal C}_n$ with a non-absorbing final state
$j$ is negligible if 
\begin{equation}
 | \tau - \langle \tau \rangle_{{\cal C}_n} | \le a \sqrt{\Theta_{{\cal C}_n}}
\label{cond}
\end{equation}
with a suitably chosen {\em accuracy parameter} $a$. For $\displaystyle \tau < 
\langle \tau \rangle_{{\cal C}_n} - a \sqrt{\Theta_{{\cal C}_n}}$ there is
not enough time to traverse all states $i_k$ of the path, whereas for
$\displaystyle \tau > \langle \tau \rangle_{{\cal C}_n} 
+ a \sqrt{\Theta_{{\cal C}_n}}$
the system will probably have already visited additional states $i_{n+1}$,
$i_{n+2}$, $\dots$ Therefore, only paths which fulfil condition (\ref{cond})
are {\em relevant}. Choosing the accuracy parameter $a \approx 3$ allows to
reconstruct about $99$\% of the
probability distribution $P(j,\tau)$ which can be
checked by means of the {\em normalization condition}
$\sum_{j=1}^N P(j,\tau) = 1$.
\par
If we are interested in the probability 
that the system takes {\em desired} paths
only, we have to take into account an even smaller number of relevant paths. 
This is simply done by restricting the summation (\ref{sechs}) to the desired
states. A similar procedure is used to determine the expected escape time
of a system from undesired states.
\par
In order to evaluate the path-related quantities introduced above, the
simulation tool EPIS ({\bf E}fficient {\bf P}ath {\bf I}ntegral {\bf
  S}imulator) has recently been developed at the University of Stuttgart
(Molini 1995). This bases on the standard path-search algorithm 
{\em depth-first} (Schildt 1990) but restricts it to the relevant paths
(see Figure 2). The advantage of this algorithm is that it uses computer
memory very efficiently and allows to calculate certain quantities like
\begin{equation}
 \langle \tau \rangle_{{\cal C}_{n+1}} = \langle \tau \rangle_{{\cal C}_n}
 + \frac{1}{w_{n+1}} \, , \quad
 \Theta_{{\cal C}_{n+1}} = \Theta_{{\cal C}_n} + \frac{1}{(w_{n+1})^2} \, ,
 \quad
 w({\cal C}_{n+1}) = w({\cal C}_n) w(i_{n+1}|i_n) 
\end{equation}
after each step. In this way, the results of previous calculations can be
utilized which again saves computer time.
\par\begin{figure}[htbp]
\unitlength0.75cm
\begin{center}
\begin{picture}(18.5,6)(0,-0.5)
\put(0.5,-0.3){\makebox(0,0){$\vdots$}}
\put(0.5,1){\makebox(0,0){3}}
\put(0.5,2){\makebox(0,0){2}}
\put(0.5,3){\makebox(0,0){1}}
\put(0.5,4){\makebox(0,0){0}}
\put(0.5,5){\makebox(0,0){$n$}}
\put(1.2,4.8){\line(-1,1){1}}
\put(1,5.5){\makebox(0,0){$Y_n$}}
\put(2,5.5){\makebox(0,0){1}}
\put(3,5.5){\makebox(0,0){2}}
\put(4,5.5){\makebox(0,0){3}}
\boldmath
\put(3,-0.5){\makebox(0,0){Step 1}}
\put(2,1){\circle{0.4}}
\put(3,1){\circle{0.4}}
\put(4,1){\circle{0.4}}
\put(2,2){\circle{0.4}}
\put(3,2){\circle{0.4}}
\put(4,2){\circle{0.4}}
\put(2,3){\circle*{0.4}}
\put(3,3){\circle{0.4}}
\put(4,3){\circle{0.4}}
\put(2,4){\circle{0.4}}
\put(3,4){\circle*{0.4}}
\put(4,4){\circle{0.4}}
\thicklines
\put(2.8,3.8){\vector(-1,-1){0.6}}
\thinlines
\put(6,5.5){\makebox(0,0){1}}
\put(7,5.5){\makebox(0,0){2}}
\put(8,5.5){\makebox(0,0){3}}
\boldmath
\put(7,-0.5){\makebox(0,0){Steps 2--4}}
\put(6,1){\circle{0.4}}
\put(7,1){\circle{0.4}}
\put(8,1){\circle{0.4}}
\put(6,2){\makebox(0,0){$\otimes$}}
\put(7,2){\makebox(0,0){$\ominus$}}
\put(8,2){\circle*{0.4}}
\put(6,3){\circle*{0.4}}
\put(7,3){\circle{0.4}}
\put(8,3){\circle{0.4}}
\put(6,4){\circle{0.4}}
\put(7,4){\circle*{0.4}}
\put(8,4){\circle{0.4}}
\thicklines
\put(6.8,3.8){\line(-1,-1){0.6}}
\put(6.25,2.85){\vector(3,-1){1.7}}
\thinlines
\put(6.05,2.7){\vector(3,-2){0.7}}
\put(6.95,2.3){\vector(-3,2){0.7}}
\put(10,5.5){\makebox(0,0){1}}
\put(11,5.5){\makebox(0,0){2}}
\put(12,5.5){\makebox(0,0){3}}
\boldmath
\put(11,-0.5){\makebox(0,0){Steps 5--11}}
\put(10,1){\makebox(0,0){$\ominus$}}
\put(11,1){\makebox(0,0){$\ominus$}}
\put(12,1){\makebox(0,0){$\otimes$}}
\put(10,2){\circle{0.4}}
\put(11,2){\circle{0.4}}
\put(12,2){\circle{0.4}}
\put(10,3){\circle{0.4}}
\put(11,3){\makebox(0,0){$\otimes$}}
\put(12,3){\circle*{0.4}}
\put(10,4){\circle{0.4}}
\put(11,4){\circle*{0.4}}
\put(12,4){\circle{0.4}}
\thicklines
\put(11.2,3.8){\vector(1,-1){0.6}}
\thinlines
\put(11.75,1.95){\vector(-3,-1){1.8}}
\put(10.15,1.25){\vector(3,1){1.7}}
\put(11.7,1.7){\vector(-3,-2){0.7}}
\put(11.2,1.2){\vector(3,2){0.8}}
\put(11.95,2.25){\vector(-3,1){1.7}}
\put(10.2,3.2){\vector(1,1){0.6}}
\put(14,5.5){\makebox(0,0){1}}
\put(15,5.5){\makebox(0,0){2}}
\put(16,5.5){\makebox(0,0){3}}
\boldmath
\put(15,-0.5){\makebox(0,0){Steps 12--14}}
\put(14,1){\circle{0.4}}
\put(15,1){\circle{0.4}}
\put(16,1){\circle{0.4}}
\put(14,2){\makebox(0,0){$\ominus$}}
\put(15,2){\circle*{0.4}}
\put(16,2){\circle{0.4}}
\put(14,3){\circle{0.4}}
\put(15,3){\circle{0.4}}
\put(16,3){\circle*{0.4}}
\put(14,4){\circle{0.4}}
\put(15,4){\circle*{0.4}}
\put(16,4){\circle{0.4}}
\thicklines
\put(15.2,3.8){\line(1,-1){0.6}}
\put(15.85,2.75){\vector(-1,-1){0.6}}
\thinlines
\put(15.75,2.95){\vector(-3,-1){1.8}}
\put(14.15,2.25){\vector(3,1){1.7}}
\unboldmath
\put(18,5.5){\makebox(0,0){$\cdots$}}
\put(18,2.5){\makebox(0,0){$\cdots$}}
\put(18,-0.5){\makebox(0,0){$\cdots$}}
\end{picture}
\end{center}
{\small Figure 2: Efficient path generation according to the modified
{\em depth-first} algorithm. Here, it is assumed that the system can 
be in one of the states $Y_n \in {\cal M} = \{1,2,3\}$ after the
$n$th transition and that it is initially in the state $Y_0 = i_0 = 2$.
The single steps of the procedure are symbolized by arrows. They correspond to
an extension of a path ${\cal C}_n$ by a new admissible state $Y_{n+1}
= i_{n+1}$ or to a removal of the last state $i_n$ if the path ${\cal C}_n$ 
cannot be further extended to a relevant path ${\cal C}_{n+1}$. Full circles
represent the states of the last generated path and are connected by
thick arrows or lines. Crossed circles (\mbox{\boldmath $\otimes$})
stand for states $Y_{n+1}$ which are not admissible since transitions from 
state $Y_{n}$ must lead to another state $Y_{n+1} \ne Y_n$. The symbol
``\mbox{\boldmath $\ominus$}'' indicates that the resulting path ${\cal
  C}_{n+1}$ turns out to be not relevant due to 
$\tau > \langle \tau \rangle_{{\cal C}_{n+1}} + a \sqrt{\Theta_{{\cal C}_{n+1}}}$.
Since then all longer paths are irrelevant, too, the procedure removes
the last state and tries to extend the remaining path ${\cal C}_n$ by another
state. If this is not possible, the state $i_n$ is also removed, etc.}
\end{figure}
The simulation tool EPIS is constructed for systematically generating
relevant paths and for calculating
\begin{itemize}
\item their occurence probabilities (\ref{form}),
\item their occurence time distributions (\ref{occtime}),
\item the occurence probability of desired paths up to time $\tau$,
\item the expected escape time from undesired states. 
\end{itemize}
Moreover, the contracted path integral solution (\ref{erg2}) 
can be compared with the
distribution $P(j,\tau)$ of states which results by a 
numerical integration of the master equation.

\section{Effective Cumulative Life-Time Distribution and Hidden State Concept}

The time-dependence of hazard rates is usually fitted to functions like the
ones belonging to the Gompertz distribution, the Weibull distribution, 
the log-normal distribution, the log-logistic distribution, the
sickle distribution, the extreme-value distribution, 
the gamma distribution, and others
(Blossfeld et al. 1986; Kalbfleisch/Prentice 1980; Elandt-Johnson/Johnson
1980; Tuma/Hannan 1984; Cox/Oakes 1984; Lancaster
1990; Courgeau/Leli\`{e}vre 1992). Since these functional
time-dependences can not always be interpreted in terms of underlying
social or psychological processes,
a new concept for an {\em interpretation} of the respective
time-dependence of transition rates is proposed
in the following. This bases on the discovery
that, if we would have sequential or parallel transitions between
{\em hidden states}, this would cause a time-dependence of the related
{\em effective} transition rate $w_{\rm eff}(\tau;j|i)$ of the {\em total}
transition process. Hidden states denote states
which are either not phenomenologically 
distinguishable from other states, not externally measurable, 
or simply not detected. They could, for example, reflect 
the psychological and mental stages preceding a concrete action,
the individual predisposition, or the internal attitude towards a certain
behavior.
\par
It should be underlined that the aim of the hidden state concept is not
to question or criticize but to supplement the powerful methods of 
classical survival analysis, which have proved their 
suitability in numerous applications. Favourable properties
of the hidden state concept are
\begin{itemize}
\item[1.] the huge number of classes of effective transition rates
which can be generated by it and which may include additional 
kinds of time-dependences,
\item[2.] the theoretical relationship between these different classes
of time-dependent transition rates,
\item[3.] the possibility of a direct interpretation of each
hidden state model.
\end{itemize}
\par
Before the hidden state concept is discussed in general
(cf. Section \ref{gen}), it will be illustrated
by two well-known cases which correspond to {\em pure birth processes} 
(cf. Section \ref{sequ}) and to the situation of {\em unobserved heterogeneity}
(cf. Section \ref{paral}), respectively. This allows to
introduce the basic ideas without complex mathematical considerations
which are unavoidable in the general case.  

\subsection{Sequential Transitions} \label{sequ}

As an example for a series of birth processes, let us face the transition
\begin{equation}
 \mbox{married} \longrightarrow \mbox{divorced}
\label{example1}
\end{equation}
and assume that it is actually a transition of the kind
\begin{equation}
 \mbox{married} \longrightarrow \mbox{married with attitude towards being divorced}
 \longrightarrow \mbox{divorced} \, .
\label{exam}
\end{equation}
It should be mentioned that this example is, of course, only a ``toy model'' 
which was chosen exclusively for illustrative reasons.
\par
In order to take into account the hidden state ``married with attitude towards
being divorced'' we can introduce the following notation of states:
\begin{eqnarray}
 (1,1) &:=& \mbox{married} \, , \nonumber \\
 (1,2) &:=& \mbox{married with attitude towards being divorced} \, , 
 \nonumber \\
 (2,1) &:=& \mbox{divorced} \, .
\end{eqnarray}
With this, example (\ref{exam}) can be written in the form
\begin{equation}
{\cal H}_2 :=  (2,1) \longleftarrow (1,2)  \longleftarrow (1,1) 
\end{equation}
or
\begin{equation}
{\cal H}_2 :=  {\bf j} \longleftarrow {\bf i}_1 
 \longleftarrow {\bf i}_0 \, ,
\label{simple}
\end{equation}
where ${\bf j} = {\bf i}_2 := (2,1)$, ${\bf i}_1 := (1,2)$, and
${\bf i}_0 := (1,1)$.
In general, the distinguished states $i \in \{1,\dots,N\}$ 
actually comprise the {\em ``hidden states''}
\begin{equation}
 {\bf i} := (i,h) \qquad \mbox{with} \qquad h \in \{1,\dots,N_i\} \, ,
\label{hidrep}
\end{equation}
where $h$ differentiates the (possibly unknown) variants of state $i$.
\par
Analogous to (\ref{lifetime}) the {\em effective cumulative 
life-time distribution}
$F^1_{\rm eff}(\tau|i)$ for the {\em total} sequence (\ref{simple})
of transitions can be defined by the probability
$P({\cal T}_2 < \tau|{\bf i}_1\leftarrow {\bf i}_0)$ that the final transition 
(i.e. the transition into state ${\bf j}$) takes place before time $\tau$
given that the preceding path was ${\bf i}_1 \longleftarrow {\bf i}_0$.
Since $P({\cal H}_2,\tau)$ is the probability to have path
${\cal H}_2$ at time $\tau$ we get
\begin{equation}
 F_{\rm eff}^1(\tau|i) = 
 P({\cal T}_2 < \tau|{\bf i}_1 \leftarrow {\bf i}_0) = P({\cal H}_2,\tau) \, .
\end{equation}
Inserting (\ref{erg1}) leads to
\begin{eqnarray}
 F^1_{\rm eff}(\tau|i)  &=&  \left( \frac{S^1(\tau|{\bf i}_0)}
 { - w({\bf i}_0) [w({\bf i}_1) - w({\bf i}_0)]}
 + \frac{S^1(\tau|{\bf i}_1)}
 { - w({\bf i}_1) [w({\bf i}_0) - w({\bf i}_1)]} \right. \nonumber \\
 & & \qquad \quad \left. + \frac{1} { w({\bf i}_1) w({\bf i}_0) }
 \right) w({\bf i}_1) w({\bf i}_0) 
\end{eqnarray}
because of $w({\bf j})= 0$, $w({\bf j}|{\bf i}_1) = w({\bf i}_1)$,
$w({\bf i}_1|{\bf i}_0) = w({\bf i}_0)$, and
$P({\bf i}_0,\tau_0) = 1$. With this we can define the {\em effective
survivor function}
\begin{equation}
 S^1_{\rm eff}(\tau|i) := 1 - F^1_{\rm eff}(\tau|i)
 = \left( \frac{S^1(\tau|{\bf i}_0)}{w({\bf i}_0)}
 - \frac{S^1(\tau|{\bf i}_1)}{w({\bf i}_1)} \right) 
 \frac{w({\bf i}_1)w({\bf i}_0)}{w({\bf i}_1) - w({\bf i}_0)} 
\label{differ1}
\end{equation}
in accordance with (\ref{accord}).
\par
In the following we try to represent this expression in the form
\begin{equation}
 S^1_{\rm eff}(\tau|i) = \exp \left[ - \!\int\limits_{\tau_0}^\tau \!
 d\tau' \, w_{\rm eff}(\tau'|i) \right]
\label{differ}
\end{equation}
(cf. (\ref{eq}))
which corresponds to the {\em detected (observed, measured)} 
transition (\ref{example1}) of the type $j \longleftarrow  i$
(where $i$ corresponds to ``married'' and $j$ to ``divorced'').
The appropriate relation for the {\em effective overall departure rate}
$w_{\rm eff}(\tau|i)$ is found by differentiation of (\ref{differ})
with respect to $\tau$. We obtain
\begin{equation}
 \frac{d}{d\tau} S^1_{\rm eff}(\tau|i) = 
 - w_{\rm eff}(\tau|i) S^1_{\rm eff}(\tau|i) \qquad \mbox{i.e.} \qquad
  w_{\rm eff}(\tau|i) = - \frac{1}{S^1_{\rm eff}(\tau|i)} 
 \frac{d}{d\tau} S_{\rm eff}^1(\tau|i) \, .
\label{formel}
\end{equation}
Inserting (\ref{differ1}) finally gives 
\begin{equation}
 w_{\rm eff}(\tau|i) = 
 \frac{S^1(\tau|{\bf i}_0) - S^1(\tau|{\bf i}_1) }
 { \displaystyle \frac{S^1(\tau|{\bf i}_0)}{w({\bf i}_0)}
 - \frac{S^1(\tau|{\bf i}_1)}{w({\bf i}_1)} } \, .
\label{all}
\end{equation}
For the {\em effective transition rate} $w_{\rm eff}(\tau;j|i)$
we find $w_{\rm eff}(\tau;j|i) = w_{\rm eff}(\tau|i)$
since there are no alternative transitions from state $i$ to state $j$.
\par
The most important fact about the relations 
for $w_{\rm eff}(\tau;j|i)$ and $w_{\rm eff}(\tau|i)$ is that these
are {\em time-dependent.}
Therefore, the hidden state concept of behavioral changes
may serve as a means for interpreting the time-dependence of hazard rates.
A comparison with empirical data, however, shows that the above model
(\ref{exam}) is still too 
simple for an explanation of the sickle curve which is found for
the hazard rate of divorces (Diekmann/Mitter 1984). 
\par
The above toy model could be improved by distinguishing different motivations
(i.e. alternative reasons) which may lead to a divorce. 
This brings us to hidden state models with parallel transitions which
correspond to cases of unobserved heterogeneity.

\subsection{Parallel Transitions} \label{paral}
 
For illustrative reasons we will discuss 
an example which stems from Kinsey 
et al. (1948) concerning the kinds of sexual activities of white males 
in the United States. It should be
mentioned that this rather old example is not intended to demonstrate 
survival analytical or statistical methods. In addition, since it is
based on a small amount of 
cross-sectional data, it only allows the determination of the most
frequent transitions. Longitudinal data would, of
course, supply more detailed information, e.g. concerning the possible
(but seemingly infrequent) transitions ``heterosexual activity'' 
$\longrightarrow$ ``bisexual activity'' or ``bisexual activity'' 
$\longrightarrow$ ``homosexual activity''. 
Therefore, the interpretation of the data will also
not be the focus of our discussion. The example was rather chosen for 
didactical reasons, since it is suitable for illuminating
\begin{itemize}
\item[1.] the hidden state model with exclusively parallel transitions,
\item[2.] possible methods for the detection and separation of hidden variables 
(here: two different kinds of ``homosexual activity'') from empirical data,
\item[3.] a simulation model basing on the master equation.
\end{itemize}
\par\begin{figure}[h]
\begin{center}
\epsfig{height=8\unitlength, angle=-90,
      bbllx=50pt, bblly=50pt, bburx=554pt, bbury=770pt,
      file=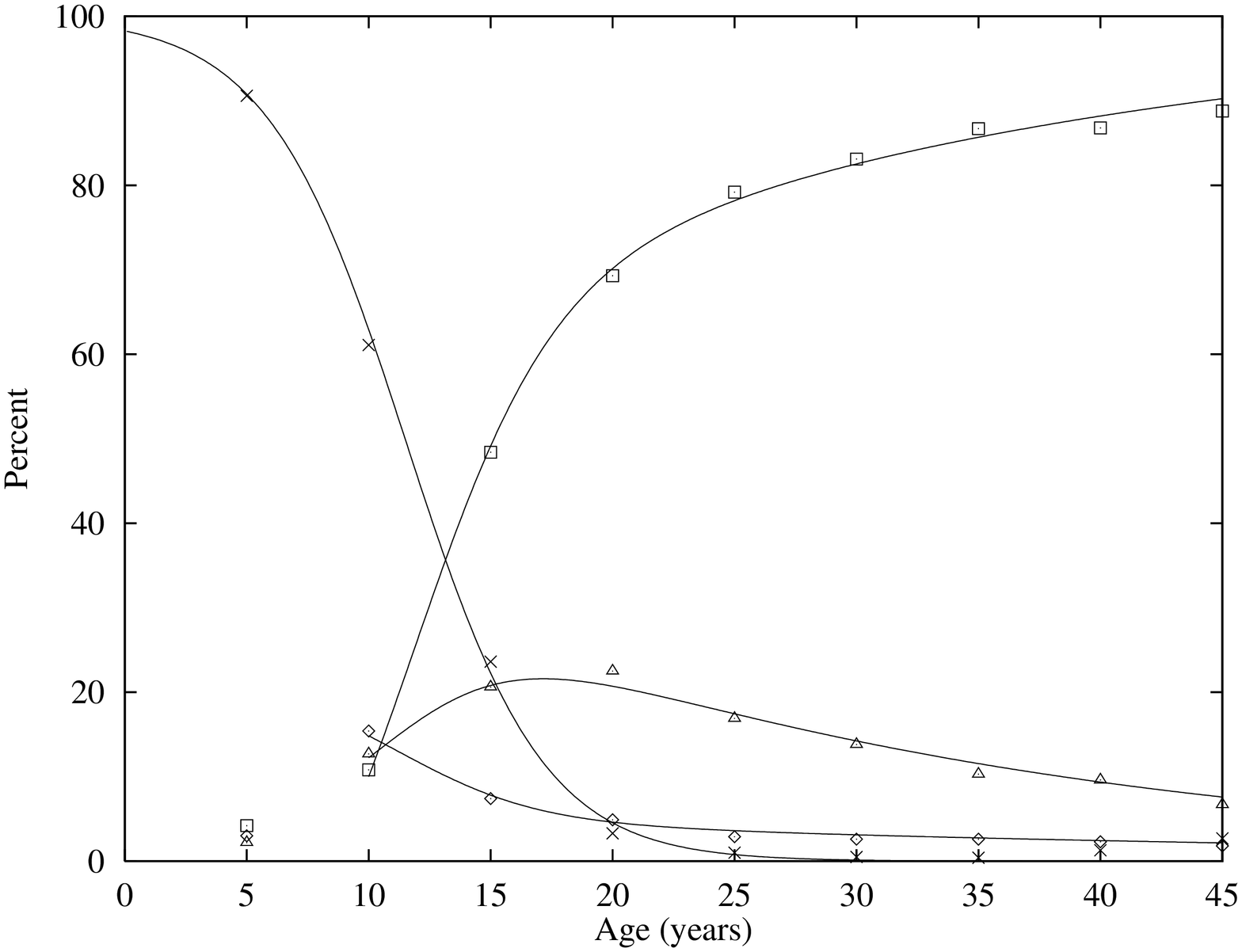}
\end{center}
{\small Figure 3: Percent of white males engaging in heterosexual activity 
($\Box$), bisexual activity ($\triangle$), homosexual activity 
($\Diamond$), 
or having no socio-sexual contacts (\mbox{\boldmath $\times$}) in dependence 
of age. Here, ``bisexual activity'' summarizes the five classes
``predominantly heterosexual, only incidentally homosexual'', 
``predominantly heterosexual, more than incidentally homosexual'',
``equally heterosexual and homosexual'', 
``predominantly homosexual, more than incidentally heterosexual'', and 
``predominantly homosexual, only incidentally heterosexual''.
The solid lines show the simulation results 
of the hidden state model which is proposed later on. Despite its simplicity
it apparently fits the data quite well.}
\end{figure}
We now come to the example by Kinsey et al. (1948).
Figure 3 shows the distribution of the different 
kinds of sexual activities (within a period of three years)
in dependence of age $\tau$. 
It turns out that the proportion 
\begin{equation}
 X_{\rm tot}(\tau) := \sum_{j=1}^3 X_j(\tau)
\end{equation}
of males with socio-sexual contacts can be approximated by the
{\em logistic equation} (Pearl 1924; Verhulst 1845)
\begin{equation}
 \frac{dX_{\rm tot}}{d\tau} = \nu X_{\rm tot}(\tau) 
 [1 - X_{\rm tot}(\tau)] \, ,
\label{logist}
\end{equation}
where $X_{\rm tot}(0 \mbox{ years}) = 0.0173$, 
$\nu = 0.36$ per year, and
\begin{eqnarray}
 X_1(\tau) &:=& \mbox{proportion of white males engaging in heterosexual activity
($j=1$)} \, ,
\nonumber \\
 X_2(\tau) &:=& \mbox{proportion of white males engaging in bisexual activity
($j=2$)} \, ,
\nonumber \\
 X_3(\tau) &:=& \mbox{proportion of white males engaging in homosexual activity
($j=3$)} \, .
\end{eqnarray}
Equation (\ref{logist}) 
could be interpreted in the way that the proportion $1 - X_{\rm tot}$
of males {\em without} socio-sexual contacts is seduced to sexual contacts
by the proportion $X_{\rm tot}$ of individuals
of about the same age {\em with} socio-sexual contacts.
\par
However, note that the logistic curve deviates from the empirical data for very
young and for old males. Whereas the proportion of males without socio-sexual
contacts should start with a value of 1 at the age of 0 years, it should
increase after an age of about 30 years due to the decrease 
of sexual opportunities (or other reasons). In any case, empirical and 
theoretical statements are very questionable before an age of 10 or 15 years.
\par
We will now investigate the proportions 
\begin{equation}
 P(j,\tau) := \frac{X_j(\tau)}{X_{\rm tot}(\tau)}
\label{perc}
\end{equation}
of white males {\em with} 
socio-sexual contacts who engage in sexual activity $j$. 
These are represented half-logarithmically in Figure 4.
$P(j,\tau)$ can also be interpreted as probability that,
in a representative sample of males with socio-sexual contacts, a randomly
picked out male engages in sexual activity $j$.
\par\begin{figure}[htbp]
\begin{center}
\epsfig{height=8\unitlength, angle=-90,
      bbllx=50pt, bblly=50pt, bburx=554pt, bbury=770pt,
      file=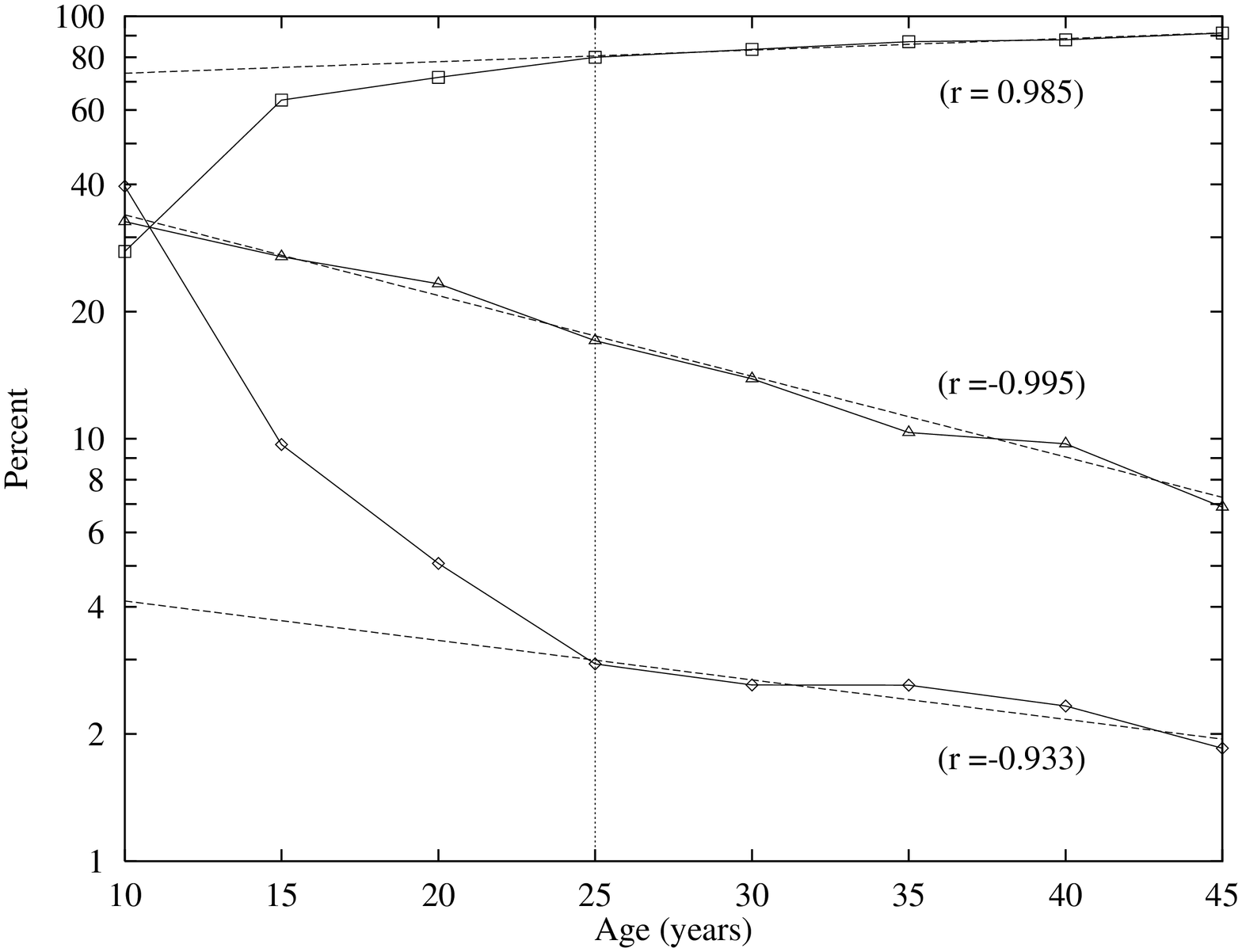}
\end{center}
{\small Figure 4: Half-logarithmic plot of the proportions of sexually 
active males engaging in heterosexual activity ($\Box$), 
bisexual activity ($\triangle$), and homosexual activity 
($\Diamond$).
The broken lines show the corresponding
linear regression curves (on the basis of the time 
period 10 years $\le \tau \le$ 45 years for bisexual activity and the time 
period 25 years $\le \tau \le$ 45 years otherwise). The numbers in brackets
are the respective correlation coefficients.}
\end{figure}
We find that heterosexual activity increases, but bisexual and homosexual
activity decrease with advancing age
which might be a consequence of an adaptation to
social norms. The half-logarithmically scaled 
curve of bisexual activity is quite well decribed by a linear
relation over the whole range of ages, whereas
the other curves behave almost linearly
after an age of 25 years. Between 10 and 25 years there seems to be a
surplus of homosexual activity connected with less heterosexual activity
than expected. This surplus decays very fast and fills the gap
of heterosexual activity. We may therefore suspect to be confronted 
with two different types of homosexual activity: 
\begin{enumerate}
\item a rather permanent
type of homosexual activity (with a half-life period of about 30 years)
which might arise from a ``homosexual predisposition'',
\item a short-lived type of ``adolescent homosexual activity''
(with a half-life period of about 2 years) which might be a substitute for
lacking opportunities for heterosexual activity.
\end{enumerate}
This interpretation is in good agreement with the findings of developmental
psychology.
\par
Summarizing the previous considerations we predominantly have the following
transitions\\[0mm]\unitlength1cm\begin{center} 
\begin{picture}(3,2)
\put(0,2){\makebox(0,0){(2,1)}}
\put(0,1){\makebox(0,0){(3,1)}}
\put(0,0){\makebox(0,0){(3,2)}}
\put(0.6,2){\vector(2,-1){1.8}}
\put(0.6,1){\vector(1,0){1.7}}
\put(0.6,0){\vector(2,1){1.8}}
\put(3,1){\makebox(0,0){(1,1)}}
\end{picture}\\[3mm]\end{center}
between the states
\begin{eqnarray}
 {\bf j} := (1,1) &:=& \mbox{``heterosexual activity''} \, , \nonumber \\
 {\bf i}_0^1 := (2,1) &:=& \mbox{``bisexual activity''} \, , \\
 {\bf i}_0^2 := (3,1) &:=& \mbox{``homosexual activity with homosexual 
 predisposition''} \, , \nonumber \\
 {\bf i}_0^3 := (3,2) &:=& \mbox{``adolescent homosexual activity,
 but heterosexual predisposition''} \, , \nonumber 
\end{eqnarray}
where we have again used the hidden state representation (\ref{hidrep}).
As Figure 3 shows, the temporal course of the proportions 
$X_j(\tau)$ can, for $\tau  \ge \tau_0 := 10$ years, 
be approximated by (\ref{logist}) and (\ref{perc}) together
with the master equation
\begin{equation}
   \frac{d}{dt} P({\bf i}',t) =
 \sum_{{\bf i} (\ne {\bf i}')} \Big[ w({\bf i}'|{\bf i}) P({\bf i},t)
 - w({\bf i}|{\bf i}') P({\bf i}',t) \Big]
\qquad ({\bf i},{\bf i}' \in \{ {\bf j},{\bf i}_0^1,{\bf i}_0^2,{\bf i}_0^3\} )
\, .
\label{mASTEr}
\end{equation}
The presented simulation results are for the parameter values 
\begin{eqnarray}
 P({\bf i}_0^1,\tau_0) = 0.33 
 \, , &\qquad & 
 w({\bf j}|{\bf i}_0^1) = 0.042 \mbox{ per year} \, , \nonumber \\
 P({\bf i}_0^2,\tau_0) = 0.05 
 \, , &\qquad & 
 w({\bf j}|{\bf i}_0^2) = 0.024 \mbox{ per year} \, , \nonumber \\
 P({\bf i}_0^3,\tau_0) = 0.35 
 \, , &\qquad & 
 w({\bf j}|{\bf i}_0^3) = 0.366 \mbox{ per year} 
\end{eqnarray}
which were determined with the {\em method of least squares}
(Elandt-Johnson/Johnson 1980; Tuma/Hannan 1984; Lancaster 1990; Helbing 1995).
Although the terms $w({\bf i}'|{\bf i})P({\bf i},t)$ 
were assumed to be negligible for ${\bf i}' \ne {\bf j}$ 
in accordance with the simplified model under consideration, the
simulation results fit the empirical data quite well.
\par
A number of alternative models with the same number of parameters
have also been tested. Most of them yielded worse correlations with 
the empirical data. Only the assumption that ``homosexual activity with
homosexual predisposition'' is replaced by ``bisexual activity'' instead of
``heterosexual activity'' produced comparable results. A better
correlation with the empirical data can, of course, always
be reached by more complicated models which include a 
greater number of parameters. However, an increase of the
number of parameters is only advisable, if this gives a considerably better
fit. Otherwise, some of the parameters will be insignificant so that the
explanatory power of at least some variables is very low.
\par
In the following we want to describe the transitions from the two ``hidden''
homosexual states ${\bf i}_0^2 = (3,1)$ and ${\bf i}_0^3 = (3,2)$ 
in an overall manner as if we would only have one homosexual state
$i = 3$. The associated effective 
transition scheme (which corresponds to
the states that were actually detected) is 
\\[0mm]\unitlength1cm\begin{center} 
\begin{picture}(2.5,1)
\put(0,1){\makebox(0,0){2}}
\put(0,0){\makebox(0,0){3}}
\put(0.4,1){\vector(4,-1){1.8}}
\put(0.4,0){\vector(4,1){1.8}}
\put(2.5,0.5){\makebox(0,0){1}}
\end{picture}\\[3mm]\end{center}
The {\em effective cumulative life-time distribution}
$F^1_{\rm eff}(\tau|i) = P({\cal T}_1 < \tau|X_0 = i)$
for the transition from homosexual to heterosexual activity is given by 
the probability that one of the paths ${\bf j} \longleftarrow {\bf i}_0^2$
or ${\bf j} \longleftarrow {\bf i}_0^3$ is taken before time $\tau$:
\begin{equation}
 F^1_{\rm eff}(\tau|i) = P({\bf j}\leftarrow {\bf i}_0^2,\tau)
 + P({\bf j}\leftarrow {\bf i}_0^3,\tau) \, .
\end{equation}
From (\ref{erg1}) we obtain
\begin{eqnarray}
 F^1_{\rm eff}(\tau|i) &=& \left( \frac{S^1(\tau|{\bf i}_0^2)}
 {w({\bf j}) - w({\bf i}_0^2)} 
 + \frac{S^1(\tau|{\bf j})}{w({\bf i}_0^2) - w({\bf j})} \right)
 w({\bf j}|{\bf i}_0^2) P({\bf i}_0^2,\tau_0) \nonumber \\
 &+& \left( \frac{S^1(\tau|{\bf i}_0^3)}
 {w({\bf j}) - w({\bf i}_0^3)} 
 + \frac{S^1(\tau|{\bf j})}{w({\bf i}_0^3) - w({\bf j})} \right)
 w({\bf j}|{\bf i}_0^3) P({\bf i}_0^3,\tau_0)
\nonumber \\
&=& [ 1 - S^1(\tau|{\bf i}_0^2)] P({\bf i}_0^2,\tau_0)
 + [ 1 - S^1(\tau|{\bf i}_0^3)] P({\bf i}_0^3,\tau_0)
\end{eqnarray}
due to $w({\bf j}|{\bf i}_0^2) = w({\bf i}_0^2)$,
$w({\bf j}|{\bf i}_0^3) = w({\bf i}_0^3)$, and $w({\bf j}) = 0$.
With $S^1_{\rm eff}(\tau|i) = 1 - F^1_{\rm eff}(\tau|i)$
we now find the {\em effective survivor function}
\begin{equation}
 S^1_{\rm eff}(\tau|i) = S^1(\tau|{\bf i}_0^2) P({\bf i}_0^2,\tau_0)
 + S^1(\tau|{\bf i}_0^3) P({\bf i}_0^3,\tau_0) \, .
\label{eseff}
\end{equation}
If we again try to express the effective survivor function in form
(\ref{differ}), we can calculate the effective overall departure rate
$w_{\rm eff}(\tau|i)$ via formula (\ref{formel}). Inserting (\ref{eseff})
provides
\begin{equation}
 w_{\rm eff}(\tau|i) = 
 \frac{w({\bf i}_0^2) S^1(\tau|{\bf i}_0^2) P({\bf i}_0^2,\tau_0)
 + w({\bf i}_0^3) S^1(\tau|{\bf i}_0^3) P({\bf i}_0^3,\tau_0)}
 {S^1(\tau|{\bf i}_0^2) P({\bf i}_0^2,\tau_0)
 + S^1(\tau|{\bf i}_0^3) P({\bf i}_0^3,\tau_0)} \, .
\label{para}
\end{equation}
This result obviously differs from expression (\ref{all})
which we obtained for the hidden state model (\ref{simple}) of the
Section \ref{sequ}.
However, since we have only transitions to {\em one} final state,
we again find $w_{\rm eff}(\tau;j|i) = w_{\rm eff}(\tau|i)$
for the {\em effective transition rates} $w_{\rm eff}(\tau;j|i)$.
\par
Note that the two hidden homosexual states are an example for the 
widespread case
of {\em unobserved heterogeneity (unmeasured heterogeneity)}
(Blossfeld et al. 1986; Elandt-Johnson/Johnson
1980; Diekmann/Mitter 1984; Tuma/Hannan 1984; Cox/Oakes 1984; Lancaster
1990; Courgeau/Leli\`{e}vre 1992). In the
general case of {\em heterogeneous populations (mixed populations)},
we have the relations
\begin{equation}
 S_{\rm eff}^1(\tau|i) = \sum_k S^1(\tau|{\bf i}_0^k) P({\bf i}_0^k,\tau_0)
\label{mix}
\end{equation}
and
\begin{equation}
 w_{\rm eff}(\tau;j|i) = w_{\rm eff}(\tau|i) = \frac{\displaystyle \sum_k
 f^1(\tau;{\bf j}|{\bf i}_0^k)P({\bf i}_0^k,\tau_0)}
 {\displaystyle \sum_k S^1(\tau|{\bf i}_0^k)P({\bf i}_0^k,\tau_0)}
\end{equation}
if there is only one final state 
${\bf j} = (j,1)$ (Courgeau/Leli\`{e}vre 1992: 44; 
Blossfeld et al. 1986: 93). Formula (\ref{mix}) describes the mixture
of survival functions in heterogeneous populations. $P({\bf i}_0^k,\tau_0)$
is called the {\em mixing distribution (compounding distribution)}, and the
state $i = i_0$ summarizing the states ${\bf i}_0^k$ can be interpreted as
{\em compound state}. In the special case that some unobserved
subpopulations (which are here reflected by $k$)
are subject to transitions but others not, we are confronted with a so-called
{\em mover-stayer model} (cf. Courgeau/Leli\`{e}vre 1992). The reason is that
the formerly mentioned subpopulations can be interpreted as {\em movers}, 
the latter ones (which are characterized by $w({\bf i}_0^k) = 0$) 
as {\em stayers}.

\subsection{The General Case} \label{gen}

An investigation of formula (\ref{all}) shows that {\em sequential} 
transitions are
related with an effective transition rate $w_{\rm eff}(\tau;j|i)$
which is zero at $\tau = \tau_0$ and {\em increases} monotonically
in the course of time up to
the limiting value $\min[w({\bf i}_0),w({\bf i}_1)]$. In contrast,
for the case (\ref{para}) of {\em parallel} transitions we find that the
corresponding effective transition rate $w_{\rm eff}(\tau;j|i)$
starts with the finite value $w_{\rm eff}(\tau_0;j|i)
= w({\bf i}_0^2) P({\bf i}_0^2,\tau_0) + w({\bf i}_0^3) P({\bf
i}_0^3,\tau_0)$ and {\em decreases} monotonically up to the lower limit 
$\min[w({\bf i}_0^2),w({\bf i}_0^3)]$. 
\par
Therefore one can conjecture
that {\em any} time-dependence of hazard rates (including sickle-shaped ones)
can be obtained or at least
approximated by a suitable combination of sequential and parallel transitions.
The general hidden state concept being necessary for this is presented in
the following and can be skipped by readers who are not interested in
the mathematical details.
\par
Again we assume that the detected states $i$ comprise a number of hidden
states ${\bf i} = (i,h)$ with $h \in \{1,2,\dots,N_{i}\}$
and $N_{i} \ge 1$. In the general case, a detected transition 
$j \longleftarrow i$ is composed of all kinds of transitions 
between the hidden states ${\bf i}_k = (i,h_k)$ comprised by state $i$ 
($h_k \in \{1,\dots,N_i\}$) before the individual or system comes the first
time into
one of the hidden states ${\bf j} = (j,h_n)$ ($h_n \in \{1,\dots,N_j\}$)
which are comprised by state $j$ (see Figure 5). 
Therefore, we will have to sum up over all paths ${\cal H}$ of the form
\begin{equation}
{\cal H}_n :=  {\bf j} \longleftarrow {\bf i}_{n-1} 
\longleftarrow \dots \longleftarrow {\bf i}_0 \qquad
\mbox{with} \quad {\bf i}_k = (i,h_k) \quad \mbox{and} \quad {\bf j} = (j,h_n)
\, , \label{pathform}
\end{equation}
where the path length $n$ is arbitrary.
For this purpose we introduce the abbreviation
\begin{equation}
 \sum_{\cal H} := \sum_{n=0}^\infty \sum_{h_n = 1}^{N_j} \sum_{{\cal H}_n}
 \qquad \mbox{with} \qquad 
 \sum_{{\cal H}_n} := \!\! 
 \sum_{h_{n-1}=1}^{N_i}
 \sum_{h_{n-2}=1 \atop (h_{n-2} \ne h_{n-1})}^{N_i} \dots
 \sum_{h_0=1 \atop (h_0 \ne h_1)}^{N_i} 
\end{equation}
similar to (\ref{sechs}).
\par\begin{figure}[htbp]
\unitlength0.75cm
\begin{center}
\begin{picture}(15,4.5)(1.8,-1.75)
\put(1.8,-0.2){\makebox(0,0){$(i,3)$}}
\put(1.8,1){\makebox(0,0){$(i,2)$}}
\put(1.8,2.2){\makebox(0,0){$(i,1)$}}
\put(1.8,-1.7){\makebox(0,0){${\bf i}_0$}}
\put(2.5,-1.7){\vector(1,0){1.6}}
\put(2.5,-0.15){\vector(3,2){1.6}}
\put(2.5,-0.1){\vector(3,4){1.6}}
\put(2.5,0.9){\vector(3,-2){1.6}}
\thicklines
\put(2.5,1.1){\vector(3,2){1.6}}
\thinlines
\put(2.5,2.1){\vector(3,-4){1.6}}
\put(2.5,2.15){\vector(3,-2){1.6}}
\put(4.8,-0.2){\makebox(0,0){$(i,3)$}}
\put(4.8,1){\makebox(0,0){$(i,2)$}}
\put(4.8,2.2){\makebox(0,0){$(i,1)$}}
\put(4.8,-1.7){\makebox(0,0){${\bf i}_1$}}
\put(5.5,-1.7){\vector(1,0){1.6}}
\put(5.5,-0.15){\vector(3,2){1.6}}
\put(5.5,-0.1){\vector(3,4){1.6}}
\put(5.5,0.9){\vector(3,-2){1.6}}
\put(5.5,1.1){\vector(3,2){1.6}}
\thicklines
\put(5.5,2.1){\vector(3,-4){1.6}}
\thinlines
\put(5.5,2.15){\vector(3,-2){1.6}}
\put(7.8,-0.2){\makebox(0,0){$(i,3)$}}
\put(7.8,1){\makebox(0,0){$(i,2)$}}
\put(7.8,2.2){\makebox(0,0){$(i,1)$}}
\put(7.8,-1.7){\makebox(0,0){${\bf i}_2$}}
\put(8.5,-1.7){\vector(1,0){1.6}}
\thicklines
\put(8.5,-0.15){\vector(3,2){1.6}}
\thinlines
\put(8.5,-0.1){\vector(3,4){1.6}}
\put(8.5,0.9){\vector(3,-2){1.6}}
\put(8.5,1.1){\vector(3,2){1.6}}
\put(8.5,2.1){\vector(3,-4){1.6}}
\put(8.5,2.15){\vector(3,-2){1.6}}
\put(10.8,1){\makebox(0,0){$\cdots$}}
\put(10.8,-1.7){\makebox(0,0){$\cdots$}}
\put(11.5,-1.7){\vector(1,0){1.6}}
\put(11.5,-0.15){\vector(3,2){1.6}}
\put(11.5,-0.1){\vector(3,4){1.6}}
\thicklines
\put(11.5,0.9){\vector(3,-2){1.6}}
\thinlines
\put(11.5,1.1){\vector(3,2){1.6}}
\put(11.5,2.1){\vector(3,-4){1.6}}
\put(11.5,2.15){\vector(3,-2){1.6}}
\put(13.8,-0.2){\makebox(0,0){$(i,3)$}}
\put(13.8,1){\makebox(0,0){$(i,2)$}}
\put(13.8,2.2){\makebox(0,0){$(i,1)$}}
\put(13.8,-1.7){\makebox(0,0){${\bf i}_{n-1}$}}
\put(14.5,-1.7){\vector(1,0){1.6}}
\thicklines
\put(14.5,-0.25){\vector(3,1){1.6}}
\thinlines
\put(14.5,-0.1){\vector(1,1){1.6}}
\put(14.5,0.9){\vector(3,-1){1.6}}
\put(14.5,1.1){\vector(3,1){1.6}}
\put(14.5,2.1){\vector(1,-1){1.6}}
\put(14.5,2.25){\vector(3,-1){1.6}}
\put(16.8,0.4){\makebox(0,0){$(j,2)$}}
\put(16.8,1.6){\makebox(0,0){$(j,1)$}}
\put(16.8,-1.7){\makebox(0,0){${\bf j}$}}
\end{picture}
\end{center}
{\small Figure 5: General hidden state model including 
parallel {\em and} sequential transitions illustrated for 
the case $N_i = 3$ and $N_j = 2$.
The possible hidden transitions leading from the detected state 
$i \in \{(i,1),(i,2),(i,3)\}$ to the detected state $j \in \{ (j,1), (j,2)\}$
are represented by arrows. The thick arrows indicate one of the numerous
paths ${\cal H}_n$ of the form (\ref{pathform}) 
which the system may take.}
\end{figure}
We are now looking for an expression which allows
to calculate the probability $F_{\rm eff}^1(\tau;j|i)$ that the system
changes, for the first time, to the detected state $j$ before time $\tau$
given that the preceding detected state was $i \ne j$
(cf. (\ref{need})). Here, we remember that
$P({\cal H}_n,\tau)$ is the probability that the path ${\cal H}_n$ is
traversed before time $\tau$, but not extended by additional states
(cf. (\ref{erg1})). Since setting $w({\bf j})=0$ guarantees that the final
state ${\bf j}$ of ${\cal H}_n$ is not left any more, $P({\cal H}_n,\tau|
w({\bf j})=0)$ is the probability that the path ${\cal H}_n$ is traversed
before time $\tau$. For this reason, $F_{\rm eff}^1(\tau;j|i)$ is given
by the sum over the probabilities $P({\cal H}_n,\tau|w({\bf j})=0)$ of
all paths ${\cal H}_n$ which have the form (\ref{pathform}):
\begin{equation}
 F^1_{\rm eff}(\tau;j|i) = 
  \frac{1}{P(i)} \sum_{{\cal H}} P({\cal H}_n,\tau|w({\bf j}) = 0) 
 = \frac{1}{P(i)} \sum_{n=0}^\infty \sum_{h_n = 1}^{N_j}
 \sum_{{\cal H}_n} P({\cal H}_n,\tau|w({\bf j}) = 0) 
  \, .
\label{of}
\end{equation}
The factor $P(i)$ takes into account that $F^1_{\rm eff}(\tau;j|i)$
is a conditional probability. It is determined by the probability
\begin{equation}
 P(i) :=  \sum_{j=1 \atop (j \ne i)}^N \sum_{n=0}^\infty \sum_{h_n = 1}^{N_j} 
 \sum_{{\cal H}_n} \lim_{\tau \rightarrow \infty} 
 P({\cal H}_n,\tau|w({\bf j}) = 0)
\end{equation}
that the system starts with the detected state $i$ and changes to another
state $j \ne i$ {\em at all}. It is obvious that the evaluation of
formula (\ref{of}) requires a computer program like EPIS.
\par
From relation (\ref{of}) we can obtain the {\em effective cumulative
life-time distribution}
\begin{equation}
 F^1_{\rm eff}(\tau|i) := \sum_{j=1 \atop (j \ne i)}^N F^1_{\rm eff}(\tau;j|i)
\end{equation}
analogous to (\ref{need2}) and confirm the desired 
{\em normalization condition} 
\begin{equation}
 \lim_{\tau \rightarrow \infty} F_{\rm eff}^1(\tau|i) = 1 \, .
\end{equation}
Moreover, we can derive the {\em effective survivor function}
\begin{equation}
 S^1_{\rm eff}(\tau|i) := 1 - F^1_{\rm eff}(\tau|i) \, ,
\end{equation}
the {\em effective probability density function}
\begin{equation}
 f^1_{\rm eff}(\tau;j|i) := \frac{d}{d\tau} F^1_{\rm eff}(\tau;j|i) \, ,
\end{equation}
and the {\em effective transition rates}
\begin{equation}
 w_{\rm eff}(\tau;j|i) := \frac{f^1_{\rm eff}(\tau;j|i)}
 {S^1_{\rm eff}(\tau|i)} 
\end{equation} 
in accordance with (\ref{three}) and (\ref{two}).
For the {\em effective overall departure rates}
\begin{equation}
 w_{\rm eff}(\tau|i) := \sum_{j=1 \atop (j\ne i)}^N w_{\rm eff}(\tau;j|i) 
\end{equation}
we again find the simple relation (\ref{formel}). 
\par
In their general form, the above formulas cannot be further evaluated. 
For illustrative reasons we will calculate the effective overall departure 
rates for the special case that there is exactly one hidden path of the
form (\ref{pathform}) which corresponds to the detected transition 
$j \longleftarrow i$. This implies that the length $n$ of the path as 
well as the states ${\bf i}_k$ and ${\bf j}$ are uniquely 
determined by $i$ and $j$,  i.e. $n = n(i,j)$ and $h_k = h_k(i,j)$
($k \in \{1,2,\dots,n\}$).
Therefore, the hidden states ${\bf i}_1, \dots , {\bf i}_{n-1}$ 
could be interpreted as {\em transient states}. 
If all overall departure rates $w_k = w(i_k)$ are pairwise different 
from each other, we find
\begin{eqnarray}
& & \hspace*{-2cm} P({\cal H}_n,\tau|w({\bf j}) = 0) \vphantom{\sum_b^b}
\nonumber \\
&=& \left( \vphantom{\int\limits_{a\atop a}^b} \right. 
 \sum_{k=0}^{n-1} \frac{S^1(\tau|{\bf i}_k)}{\displaystyle [0 - w({\bf i}_k)]
 \prod_{l=0 \atop (l\ne k)}^{n-1} [w({\bf i}_l) - w({\bf i}_k)] } 
 + \frac{1}{\displaystyle 
 \prod_{l=0 \atop (l \ne n)}^{n} [w({\bf i}_l) - 0 ] }
 \left. \vphantom{\int\limits_{a\atop a}^b} \right)
 w({\cal H}_n)P({\bf i}_0,\tau_0) 
\label{hund}
\end{eqnarray}
and, due to $w({\bf i}_k) > 0$, 
\begin{equation}
 \lim_{\tau \rightarrow \infty}
 P({\cal H}_n,\tau|w({\bf j}) = 0)
 = \frac{1}{\displaystyle 
 \prod_{l=0}^{n-1} w({\bf i}_l) }
 w({\cal H}_n)P({\bf i}_0,\tau_0) \, .
\label{hundeins}
\end{equation} 
This finally yields 
\begin{equation}
F_{\rm eff}^1(\tau|i) 
= \frac{\displaystyle 
 \left( \vphantom{\int\limits_{a\atop a}^b} \right.
 \sum_{k=0}^{n-1} \frac{S^1(\tau|{\bf i}_k)}{\displaystyle - w({\bf i}_k)
 \prod_{l=0 \atop (l\ne k)}^{n-1} [w({\bf i}_l) - w({\bf i}_k)] } 
 + \frac{1}{\displaystyle 
 \prod_{l=0}^{n-1} w({\bf i}_l) }
 \left. \vphantom{\int\limits_{a\atop a}^b} \right)
 \sum_{j=1 \atop (j \ne i)}^N w({\cal H}_n)P({\bf i}_0,\tau_0) }
 {\displaystyle \frac{1}{\displaystyle 
 \prod_{l=0}^{n-1} w({\bf i}_l) }
 \sum_{j=1 \atop ( j \ne i)}^N w({\cal H}_n)P({\bf i}_0,\tau_0) }
\end{equation}
(which can be further simplified) and
\begin{equation}
 S_{\rm eff}^1(\tau|i) 
 = \left( \prod_{l=0}^{n-1} w({\bf i}_l) \right)
 \sum_{k=0}^{n-1} \frac{S^1(\tau|{\bf i}_k)}{\displaystyle w({\bf i}_k)
 \prod_{l=0 \atop (l\ne k)}^{n-1} [w({\bf i}_l) - w({\bf i}_k)] } \, .
\label{also}
\end{equation}
$S_{\rm eff}^1(\tau|i)$ is obviously a direct generalization of $S^1(\tau|i)$
since, for $n=1$, we find
\begin{equation}
 S_{\rm eff}^1(\tau|i) = w({\bf i}_0) \frac{S^1(\tau|{\bf i}_0)}{w({\bf i}_0)}
 = S^1(\tau|{\bf i}_0) = S^1(\tau|i) \, .
\end{equation}
Finally, we arrive at the desired result
\begin{equation}
 w_{\rm eff}(\tau|i) = \frac{\displaystyle
 \sum_{k=0}^{n-1} \frac{S^1(\tau|{\bf i}_k)}{\displaystyle 
 \prod_{l=0 \atop (l\ne k)}^{n-1} [w({\bf i}_l) - w({\bf i}_k)] } }
 {\displaystyle 
 \sum_{k=0}^{n-1} \frac{S^1(\tau|{\bf i}_k)}{\displaystyle w({\bf i}_k)
 \prod_{l=0 \atop (l\ne k)}^{n-1} [w({\bf i}_l) - w({\bf i}_k)] } } 
\label{finar}
\end{equation}
due to 
\begin{equation}
 \frac{d}{d\tau} S^1(\tau|{\bf i}_k) = - w({\bf i}_k) S^1(\tau|{\bf i}_k) \, .
\end{equation}
Although the special formula (\ref{finar})
is restricted to hidden state models consisting
of sequences of $n$ birth processes, 
it is already much more complicated than (\ref{all}).

\section{Summary and Outlook}

In this paper we showed that, in the Markov case, survival analysis is related
with the master equation which describes the temporal evolution of the
distribution of states. The simulation of the master equation is
a powerful technique for the investigation of various stochastically 
behaving systems in physics, chemistry, biology, and the social sciences.
It is particularly suited for scenario techniques,
since the numerical integration of the master equation is normally
quite simple and fast.
\par
However, some interesting questions related with the longitudinal time series
of stochastic processes cannot be answered by means of the master equation.
Whereas the microsimulation of sample time series can be done with the
Monte-Carlo technique, it is rather inefficient for the numerical
determination of quantities related with the sequencing of time series.
Therefore, the simulation tool EPIS has recently been developed at the
University of Stuttgart. It facilitates the generation of relevant paths and
the evaluation of formulas which we were able to derive for path-related
quantities. This includes the occurence probabilities and
occurence time distributions of paths, 
or the expected escape time from undesired states.
\par
Finally, the formula for the occurence probabilities of paths allowed
to develop a hidden state concept of behavioral changes
which can serve as a means for interpreting the respective time-dependence
of hazard rates.
Starting from a certain hidden state model, it is possible to derive the
corresponding effective transition rates which can be compared
with the time-dependence of the empirically obtained hazard rates. 
The different steps which are necessary for determining a suitable hidden state
model and the corresponding parameter values (including the hazard rates)
were illustrated by a concrete example.
\par
Present research focuses on the investigation of the following questions:
\begin{enumerate}
\item Which kinds of time-dependences can be interpreted or approximated
in terms of a hidden state model?
\item Does the time-dependence of hazard rates determine the corresponding
hidden state model in a unique way?
\item If not, which are the transition schemes of the alternative
hidden state models and which of them is the simplest or most plausible
one in terms of a sociological or psychological interpretation?
\end{enumerate}

\subsection*{Acknowledgements}

The author wants to thank Andreas Diekmann, Ulrich Mueller,
Klaus Troitzsch, and Nigel Gilbert for their inspiring comments.
Moreover he is grateful to Volker Sommer for drawing his attention
to the example from the Kinsey report.

\section*{References}

{\small Binder, K. (1979) {\em Monte Carlo Methods in Statistical Physics}.
Berlin: Springer.\\[1.5mm]
Blossfeld, H.-P., A. Hamerle, and K. U. Mayer (1986)
{\em Ereignisanalyse}. Frankfurt: Campus.\\[1.5mm]
Chiang, C. L. (1968) {\em Introduction to Stochastic Processes in
Biostatistics}. New York: Wiley.\\[1.5mm]
Courgeau, D. and \'{E}. Leli\`{e}vre (1992) {\em Event History Analysis
in Demography}. Oxford: Clarendon Press.\\[1.5mm]
Cox, D. R. (1975) ``Partial likelihood'', {\em Biometrica} {\bf 62},
269--276.\\[1.5mm]
Cox, D. R. and D. Oakes (1984) {\em Analysis of Survival Data}.
London: Chapman and Hall.\\[1.5mm]
Diekmann, A. and P. Mitter, eds. (1984) {\em Stochastic Modelling of Social 
Processes}, especially the contribution by A. Diekmann and
P. Mitter: ``A comparison of the `sickle function' with alternative
stochastic models of divorce rates'', pp. 123--153.  Orlando: Academic
Press.\\[1.5mm]
Elandt-Johnson, R. C. and N. L. Johnson (1980) {\em Survival Models
and Data Analysis}. New York: Wiley.\\[1.5mm]
Empacher, N. (1992) {\em Die Wegintegrall\"osung der Mastergleichung}.
University of Stuttgart: PhD thesis.\\[1.5mm]
Haken, H. (1983) {\em Synergetics,} 3rd edition. Berlin: Springer.\\[1.5mm]
Helbing, D. (1994) ``A contracted path integral solution of the discrete
master equation''. {\em Physics Letters A} {\bf 195}, 128--134.\\[1.5mm]
Helbing, D. (1995) {\em Quantitative Sociodynamics. Stochastic Methods and
Models of Social Interaction Processes}. Dordrecht: Kluwer Academic.\\[1.5mm]
Helbing, D. and Molini, R. (1995) ``Occurence probabilities of stochastic
paths'', {\em Physics Letters A} (submitted).\\[1.5mm]
Kalbfleisch, J. D. and R. L. Prentice (1980) {\em The Statistical 
Analysis of Failure Time Data}. New York: Wiley.\\[1.5mm]
Kinsey, A. C., W. B. Pomeroy, and C. E. Martin (1948) {\em Sexual Behavior
in the Human Male.} Philadelphia: Saunders.\\[1.5mm]
Lancaster, T. (1990) {\em The Econometric Analysis of Transition Data}.
Cambridge: Cambridge University Press.\\[1.5mm] 
Molini, R. (1995) {\em Algorithmen und Anwendungen zur
Pfadsummenl\"osungsmethode der Masterequation}. University of Stuttgart:
Master's thesis.\\[1.5mm]
Pearl, R. (1924) {\em Studies in Human Biology}. Baltimore: Williams and 
Wilkins.\\[1.5mm]
Press, W. H., S. A. Teukolsky, W. T. Vetterling, and B. P. Flannery
(1992) {\em Numerical Recipes in C. The Art of Scientific Computing,}
2nd edition. Cambridge: Cambridge Univerisity Press.\\[1.5mm]
Schildt, H. (1990) {\em C: The Complete Reference}. Berkeley: McGraw
Hill.\\[1.5mm]
Stewart, I. and D. Tall (1983) {\em Complex Analysis}. Cambridge:
Cambridge University Press.\\[1.5mm]
Troitsch, K. G. (1990) {\em Modellbildung und Simulation in den
Sozialwissenschaften}. Opladen: Westdeutscher Verlag.\\[1.5mm]
Tuma, N. B. and M. T. Hannan (1984) {\em Social Dynamics.
Models and Methods.} Orlando: Academic Press.\\[1.5mm]
Verhulst, P.-F. (1845) ``Recherches math\'{e}matiques sur la loi
d'accroissement de la population''. {\em Nuoveaux M\'{e}moires de
l'Acad\'{e}mie Royale des Sciences et Belles-Lettres de Bruxelles} {\bf 18},
1--41.\\[1.5mm]
Weidlich, W. (1991) ``Physics and social science---The approach of
synergetics''. {\em Physics Reports} {\bf 204}, 1--163.\\[1.5mm]
Weidlich, W. and G. Haag (1983) {\em Concepts and Models of a
Quantitative Sociology. The Dynamics of Interacting Populations}. 
Berlin: Springer.}
\end{document}